\documentclass[journal,10pt]{IEEEtran}

\usepackage{cite}
\usepackage{mathrsfs}
\usepackage{amsfonts}
\usepackage{ifpdf}
\usepackage{cite}
\usepackage{stfloats}
\ifCLASSINFOpdf
\usepackage[pdftex]{graphicx}
\else \fi
\usepackage[cmex10]{amsmath}
\usepackage{array}
\usepackage{mdwmath}
\usepackage{mdwtab}
\usepackage{eqparbox}
\usepackage[tight,footnotesize]{subfigure}
\usepackage{algorithmic}
\usepackage{algorithm}
\usepackage{amsmath}
\usepackage{epsfig}
\usepackage{ae}
\usepackage{amssymb}
\usepackage{times}
\usepackage{amsmath}   
\usepackage{booktabs}  
\usepackage{threeparttable} 
\usepackage{multirow}       
\usepackage{xcolor}
\usepackage{paralist}
\usepackage{cases}
\newtheorem{theorem}{Theorem}
\newtheorem{lemma}{Lemma}
\newtheorem{corollary}{Corollary}

\newtheorem{remark}{Remark}

\begin{document}

\title{On Performance Modeling for MANETs under General Limited Buffer Constraint}

\author{Jia~Liu, Yang~Xu, Yulong~Shen, Xiaohong~Jiang, and~Tarik~Taleb
}

\maketitle

\begin{abstract}
Understanding the real achievable performance of mobile ad hoc networks (MANETs) under practical network constraints is of great importance for their applications in future highly heterogeneous wireless network environments. This paper explores, for the first time, the performance modeling for MANETs under a general limited buffer constraint, where each network node maintains a limited source buffer of size $B_s$ to store its locally generated packets and also a limited shared relay buffer of size $B_r$ to store relay packets for other nodes. Based on the Queuing theory and birth-death chain theory, we first develop a general theoretical framework to fully depict the source/relay buffer occupancy process in such a MANET, which applies to any distributed MAC protocol and any mobility model that leads to the uniform distribution of nodes' locations in steady state. With the help of this framework, we then derive the exact expressions of several key network performance metrics, including achievable throughput, throughput capacity, and expected end-to-end delay. We further conduct case studies under two network scenarios and provide the corresponding theoretical/simulation results to demonstrate the application as well as the efficiency of our theoretical framework. Finally, we present extensive numerical results to illustrate the impacts of buffer constraint on the performance of a buffer-limited MANET. 
\end{abstract}


\IEEEpeerreviewmaketitle

\section{Introduction}
The mobile ad hoc networks (MANETs), a class of self-autonomous and flexible wireless networks, are highly appealing for lots of critical applications, like disaster relief, battlefield communications, D2D communications for traffic offloading, and coverage extension in future 5G cellular networks \cite{Ramanathan_CM02,Tehrani_magzine14,Shariatmadari_CM15}. In particular, the applications of MANETs in vehicle-to-vehicle communications, i.e., the vehicular ad hoc networks (VANETs) have attracted considerable academic attention recently as a promising solution to improving safety and driving experience \cite{Hartenstein_CM08,Luo_TVT16}. Motivated by these, understanding the fundamental performance limits of MANETs is of great importance to facilitate the application and commercialization of such networks \cite{Andrews_CM08,Goldsmith_CM11}. By now, extensive works have been devoted to the performance study of MANETs, which can be roughly classified into two categories, the ones with the consideration of practical limited buffer constraint and the ones without such consideration. 

Regarding the performance study for MANETs without the buffer constraint, Grossglauser and Tse \cite{Grossglauser_TON02} first explored the capacity scaling law, i.e., how the per node throughput scales in the order sense as the number of network nodes increases, and demonstrated that with the help of node mobility a $\Theta(1)$ per node throughput is achievable in such networks. Later, Neely \emph{et al.} \cite{Neely_IT05} studied the delay-throughput tradeoff issue in a MANET under the independent and identically distributed (i.i.d) mobility model and showed that achievable delay-to-throughput ratio is lower bounded as $delay/throughput \geq O(n)$ (where $n$ is the number of network nodes). Gamal \emph{et al.} \cite{Gamal_IT06} then explored the delay-throughput tradeoff under a symmetric random walk mobility model, and showed that a $\Theta(n\log n)$ average packet delay is incurred to achieve the $\Theta(1)$ per node throughput there. Sharma \emph{et al.} \cite{Sharma_TON07} further studied the delay-throughput tradeoff under a general and unified mobility model, and revealed that there exists a critical value of delay below which the node mobility is not helpful for capacity improvement. Recently, Wang \emph{et al.} explored the throughput and delay performance for MANETs with multicast traffic in \cite{Wang_TON11,Wang_INFOCOM11}, and further conducted the network performance comparison between the unicast and multicast MANETs in \cite{Qin_TON15}. Those results indicate that the mobility can significantly decrease the multicast gain on per node capacity and delay, and thus weaken the distinction between the two traffic models. 

While the above works represent a significant progress in the performance study of MANETs, in a practical MANET, however, the buffer size of a mobile node is usually limited due to both its storage limitation and computing limitation. Thus, understanding the real achievable performance of MANETs under the practical limited buffer constraint is of more importance for the design and performance optimization of such networks. By now, some initial results have been reported on the performance study of MANETs under buffer constraint \cite{Herdtner_INFOCOM05,Gao_15,Liu_TWC15,Liu_ADHOC15}. Specifically, Herdtner and Chong \cite{Herdtner_INFOCOM05} explored the throughput-storage tradeoff in MANETs and showed that the throughput capacity under the relay buffer constraint scales as $O(\sqrt{b/n})$ (where $b$ is the relay buffer size of a node). Gao \emph{et al.} \cite{Gao_15} considered a MANET with limited source buffer in each node, and derived the corresponding cumulative distribution function of the source delay. Recently, the throughput and delay performance of MANETs are further explored under the scenarios where each node is equipped with an infinite source buffer and a shared limited relay buffer \cite{Liu_TWC15,Liu_ADHOC15}. 

\subsection{Motivation}
The motivation of our study is to take a step forward in the practical performance modeling for MANETs. In particular, this paper focuses on a practical MANET where each network node maintains a limited source buffer of size $B_s$ to store its locally generated packets and also a limited shared relay buffer of size $B_r$ to store relay packets for all other nodes. This buffer constraint is general in the sense that it covers all the buffer constraint assumptions adopted in available works as special cases, like the infinite buffer assumption \cite{Grossglauser_TON02,Neely_IT05,Gamal_IT06,Sharma_TON07,Wang_TON11,Wang_INFOCOM11,Qin_TON15} ($B_s \to \infty$, $B_r \to \infty$), limited source buffer assumption \cite{Gao_15} ($0 \leq B_s<\infty,B_r \to \infty$), and limited relay buffer assumption \cite{Herdtner_INFOCOM05,Liu_TWC15,Liu_ADHOC15} ($B_s \to \infty,0 \leq B_r<\infty$). It should be pointed out that compared with the previous works \cite{Liu_TWC15,Liu_ADHOC15} where packet loss never occurs, under the general limited-buffer scenario packet loss is inevitable, which makes deriving achievable throughput a new challenging and significant problem, and the impacts of feedback mechanism on network performance worthy of study. To the best of our knowledge, this paper represents the first attempt on the exact performance modeling for MANETs under general limited-buffer constraint.

\subsection{Our Contributions}
The main contributions of this study are summarized as follows:

\begin{itemize}
\item
Based on the Queuing theory and birth-death chain theory, we first develop a general theoretical framework to fully depict the source/relay buffer occupancy process in a MANET with the general limited-buffer constraint, which applies to any distributed MAC protocol and any mobility model that leads to the uniform distribution of nodes' locations in steady state.

\item
With the help of this framework, we then derive the exact expressions of several key network performance metrics, including achievable throughput, throughput capacity, and expected end-to-end (E2E) delay. We also provide the related theoretical analysis to reveal the fundamental network performance trend as the buffer size increases.

\item
We further conduct case studies under two network scenarios and provide the corresponding theoretical/simulation results to demonstrate the efficiency and application of our theoretical framework. Finally, we present extensive numerical results to illustrate the impacts of buffer constraint on network performance and our theoretical findings.

\end{itemize}

The remainder of this paper is organized as follows. Section~\ref{section:preliminaries} introduces the preliminaries involved in this paper. We analyze the buffer occupancy processes in Section~\ref{section:framework} and derive the exact expressions for throughput, throughput capacity and expected E2E delay in Section~\ref{section:performance}. The case studies and simulation results are presented in Section~\ref{section:case_studies}. The numerical results and corresponding discussions are provided in Section~\ref{section:numerical_results}. Finally, we conclude this paper in Section~\ref{section:conclusion}.

\section{Preliminaries} \label{section:preliminaries}

In this section, we introduce the system models, the general limited buffer constraint, the routing scheme and performance metrics involved in this study.

\subsection{System Models}

\emph{Network Model}: We consider a time-slotted MANET, which consists of $n$ nodes randomly moving in a torus network area following a ``uniform type'' mobility model. With such mobility model, the location process of a node is stationary and ergodic with stationary distribution uniform on the network area, and the trajectories of different nodes are independent and identically distributed. It is notable that such ``uniform type'' mobility model covers many typical mobility models as special cases, like the i.i.d model \cite{Neely_IT05}, random walk model \cite{Gamal_IT06}, and random direction model \cite{Nain_INFOCOM05}.

\emph{Traffic Model}: We consider that there are $n$ unicast traffic flows in the network, each node is the source of one traffic flow and also the destination of another traffic flow. More formally, let $\varphi(i)$ denote the destination node of the traffic flow originated from node $i$, then the source-destination pairs are matched in a way that the sequence $\{\varphi(1),\varphi(2),\cdots,\varphi(n)\}$ is just a derangement of the set of nodes $\{1,2,\cdots,n\}$. This traffic model is widely adopted in other studies on the performance analysis of MANETs \cite{Grossglauser_TON02,Neely_IT05,Sharma_TON07}. Two typical examples are $\varphi(1)=2$, $\varphi(2)=1$, $\varphi(3)=4$, $\varphi(4)=3$, $\cdots$, $\varphi(n-1)=n$, $\varphi(n)=n-1$ ($n$ is even), and $\varphi(1)=2$, $\varphi(2)=3$, $\cdots$, $\varphi(n)=1$. The packet generating process at each node is assumed to be a Bernoulli process with mean rate $\lambda_s^{\text{\scriptsize{+}}}$, so that with probability $\lambda_s^{\text{\scriptsize{+}}}$ a new packet is generated in each time slot. During a time slot the total amount of data that can be transmitted from a transmitter to its corresponding receiver is fixed and normalized to one packet.

\subsection{General Buffer Constraint}

\begin{figure}[!t]
\centering
\includegraphics[width=3.0in]{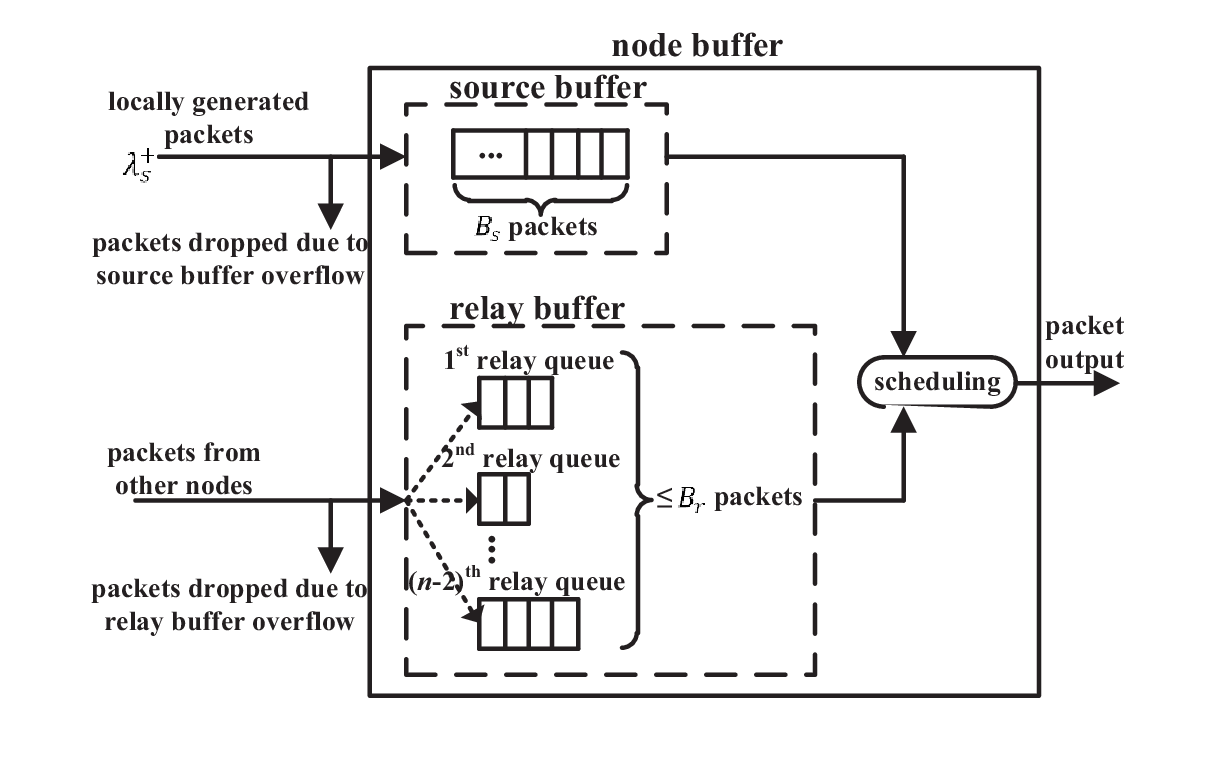} 
\caption{Illustration of the general limited buffer constraint.}
\label{fig:buffer_constraint}
\end{figure}

As illustrated in Fig.~\ref{fig:buffer_constraint}, we consider a general limited buffer constraint, where a node is equipped with a limited source buffer of size $B_s$ and a limited relay buffer of size $B_r$. The source buffer is for storing the packets of its own flow (locally generated packets) and works as a FIFO (first-in-first-out) source queue \cite{Robertazzi_BOOK12}, while the relay buffer is for storing packets of all other $n-2$ flows and works as $n-2$ FIFO virtual relay queues (one queue per flow). When a packet of other flows arrives and the relay buffer is not full, the corresponding relay queue is dynamically allocated a buffer space; once a head-of-line (HoL) packet departs from its relay queue, this relay queue releases a buffer space to the common relay buffer. It is notable that the considered limited buffer constraint is general in the sense it covers all the buffer constraint assumptions adopted in the available works as special cases.

\subsection{Two-Hop Relay Routing without/with Feedback}
Regarding the packet delivery scheme, we consider the two-hop relay (2HR) routing protocol. The 2HR scheme is simple yet efficient, and has been widely adopted in available studies on the performance modeling of MANETs \cite{Grossglauser_TON02,Neely_IT05}. In addition to the conventional 2HR scheme without feedback, we also consider the 2HR scheme with feedback, which avoids packet loss caused by relay buffer overflow and thus can support the more efficient operation of buffer-limited MANETs.

Without loss of generality, we focus on a tagged flow and denote its source node and destination node as $\mathcal{S}$ and $\mathcal{D}$ respectively. Once $\mathcal{S}$ gets access to wireless channel at the beginning of a time slot, it executes the 2HR scheme without/with feedback as follows.

\begin{enumerate}[1)]

\item
(\textbf{Source-to-Destination}) 

If $\mathcal{D}$ is within the transmission range of $\mathcal{S}$, $\mathcal{S}$ executes the Source-to-Destination operation. If the source queue of $\mathcal{S}$ is not empty, $\mathcal{S}$ transmits the HoL packet to $\mathcal{D}$; else $\mathcal{S}$ remains idle. 

\item
If $\mathcal{D}$ is not within the transmission range of $\mathcal{S}$, $\mathcal{S}$ randomly designates one of the nodes (say $\mathcal{R}$) within its transmission range as its receiver, and chooses one of the following two operations with equal probability.

\begin{itemize}

\item
(\textbf{Source-to-Relay}) 

\emph{Without feedback}: If the source queue of $\mathcal{S}$ is not empty, $\mathcal{S}$ transmits the HoL packet to $\mathcal{R}$; else $\mathcal{S}$ remains idle.

\emph{With feedback}: $\mathcal{R}$ sends a feedback to $\mathcal{S}$ to indicate whether its relay buffer is full or not. If the relay buffer of $\mathcal{R}$ is not full, $\mathcal{S}$ executes the same operation as that without feedback; else $\mathcal{S}$ remains idle.

\item
(\textbf{Relay-to-Destination}) 

In this operation, $\mathcal{S}$ serves as the relay node forwarding packets to $\mathcal{R}$, and $\mathcal{R}$ is the destination of packets forwarded from $\mathcal{S}$. If $\mathcal{S}$ has packet(s) in the corresponding relay queue for $\mathcal{R}$, $\mathcal{S}$ sends the HoL packet of this queue to $\mathcal{R}$; else $\mathcal{S}$ remains idle.

\end{itemize}

\end{enumerate}

We let $p_{sd}$, $p_{sr}$ and $p_{rd}$ denote the probabilities that a node gets the chance to execute the Source-to-Destination, Source-to-Relay, and Relay-to-Destination operations, respectively\footnote{It should be noted that a node getting the chance to execute one operation in a time slot doesn't mean that it will conduct a transmission in this time slot.}. It is worth noting that these probabilities are determined by the specific MANET scenario and will be regarded as known quantities in the following two sections, where the performance modeling is developed for a general MANET based on the basic system models mentioned above. The evaluations of $p_{sd}$, $p_{sr}$ and $p_{rd}$ will be shown in the case studies of Section~\ref{section:case_studies}.

\subsection{Performance Metrics}
The performance metrics involved in this paper are defined as follows.

\textbf{Throughput}: The \emph{throughput} $T$ of 
a flow (in units of packets per slot) is defined as the time-average number of packets that can be delivered from its source to its destination. 

\textbf{Throughput Capacity}: For the homogeneous finite buffer network scenario considered in this paper, the network level \emph{throughput capacity} $T_c$ can be defined by the maximal achievable per flow throughput, i.e., $T_c=\max\limits_{\lambda_s^{\text{\scriptsize{+}}}\in (0,1]} T$.

\textbf{End-to-end Delay}: The \emph{end-to-end delay} $D$ of a packet\footnote{Notice that for the calculation of end-to-end delay, we only focus on the packets that have been successfully delivered to their destinations.} (in units of slots) is defined as the time it takes the packet to reach its destination after it is generated by its source, and we use $\mathbb{E}\{D\}$ to denote the expectation of $D$.

\section{Buffer Occupancy Process Analysis} \label{section:framework}

In this section, we conduct the occupancy process analysis for both the source and relay buffers to determine their occupancy state distributions (OSDs), which will further help us to derive the exact expressions of the performance metrics $T$, $T_c$ and $\mathbb{E}\{D\}$. Without loss of generality, we focus on a tagged node $\mathcal{S}$, and consider the scenarios without and with feedback, respectively.

\subsection{OSDs Analysis under the Scenario without Feedback} \label{subsection:OSD_nofeedback}

\subsubsection{OSD of Source Buffer}
\begin{figure}[!t]
\centering
\includegraphics[width=3.3in]{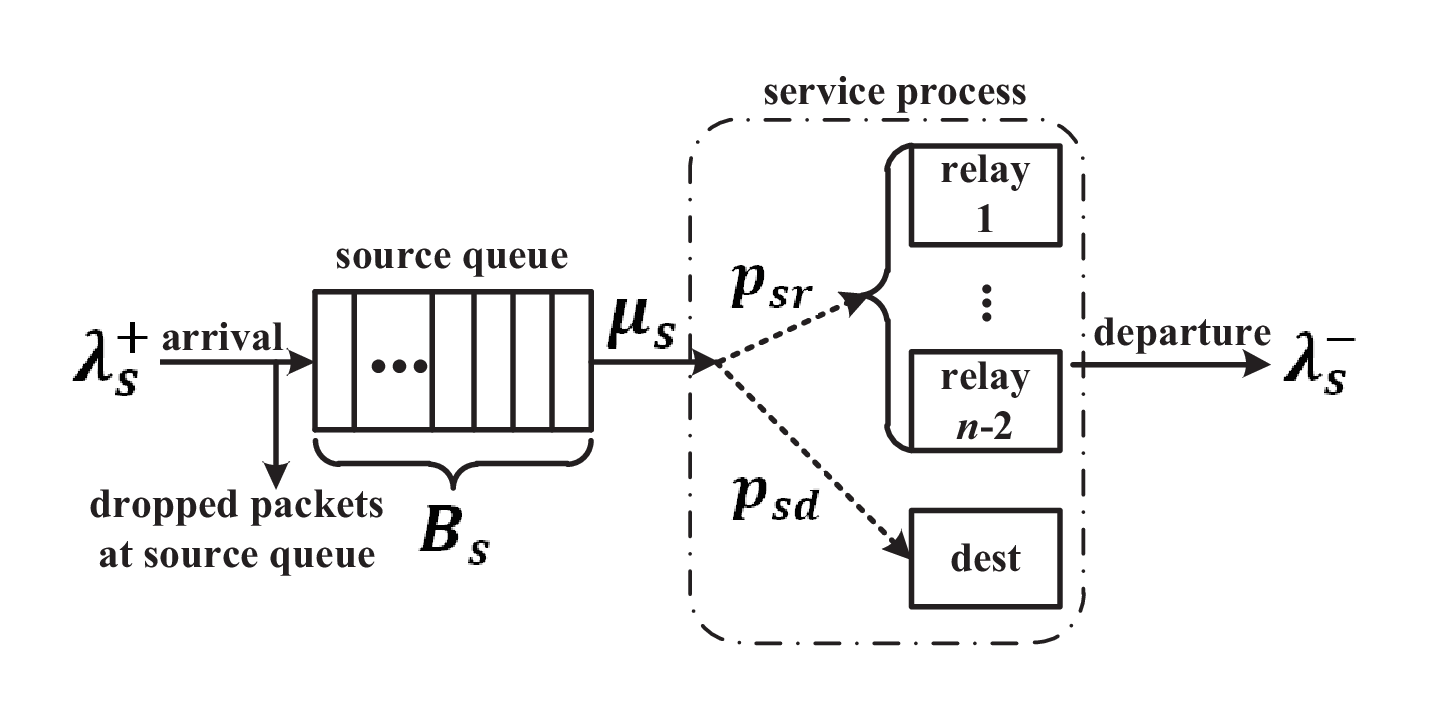} 
\caption{Bernoulli/Bernoulli/1/$B_s$ queuing model for source buffer.}
\label{fig:source_queue}
\end{figure}

Regarding the source buffer of node $\mathcal{S}$, since in every time slot a new packet is generated with probability $\lambda_s^{\text{\scriptsize +}}$ and a service opportunity arises with probability $\mu_s$ being determined as
\begin{equation}
\mu_s=p_{sd}+p_{sr}, \label{eq:mu_s_NF}
\end{equation}
the occupancy process of source buffer can be modeled by a B/B/1/$B_s$ queue as illustrated in Fig.~\ref{fig:source_queue}.

Let $\pi_s(i)$ denote the probability that there are $i$ packets occupying the source buffer in the stationary state, then the stationary OSD of the source buffer $\mathbf{\Pi}_s=\left[\pi_s(0),\pi_s(1),\cdots,\pi_s(B_s) \right]$ can be determined as \cite{Daduna_BOOK01}
\begin{equation*}
\pi_s(i)=\left\{
\begin{aligned}
&\frac{1}{1-\lambda_s^{\text{\scriptsize +}}} H^{-1}, & &i=0 \\
&\frac{1}{1-\lambda_s^{\text{\scriptsize +}}} \frac{\tau^i}{1-\mu_s} H^{-1}, & &1\leq i \leq B_s
\end{aligned}
\right.
\end{equation*}
where 
\begin{equation}
\tau=\frac{\lambda_s^{\text{\scriptsize +}}(1-\mu_s)}{\mu_s(1-\lambda_s^{\text{\scriptsize +}})}, \label{eq:tau}
\end{equation}
and $H$ is the normalization constant. Notice that $\mathbf{\Pi}_s \mathbf{\cdot 1}=1$, where $\mathbf{1}$ is a column vector of size $(B_s+1)\times 1$ with all elements being $1$, we have
\begin{equation}
\pi_s(i)=\left\{
\begin{aligned}
&\frac{\mu_s-\lambda_s^{\text{\scriptsize +}}}{\mu_s-\lambda_s^{\text{\scriptsize +}} \cdot \tau^{B_s}}, & &i=0 \\
&\frac{\mu_s-\lambda_s^{\text{\scriptsize +}}}{\mu_s-\lambda_s^{\text{\scriptsize +}} \cdot \tau^{B_s}} \frac{1}{1-\mu_s} \tau^i. & &1\leq i \leq B_s
\end{aligned}
\right. 
\label{eq:OSD_source}
\end{equation}

\subsubsection{OSD of Relay Buffer}
We continue to analyze the occupancy process of the relay buffer in $\mathcal{S}$. Let $X_t$ denote the number of packets in the relay buffer at time slot $t$, then the occupancy process of the relay buffer can be regarded as a stochastic process $\{X_t,t=0,1,2,\cdots\}$ on state space $\{0,1,\cdots,B_r\}$. Notice that when $\mathcal{S}$ serves as a relay in a time slot, the Source-to-Relay transmission and Relay-to-Destination transmission will not happen simultaneously. Thus, suppose that the relay buffer is at state $i$ in the current time slot, only one of the following transition scenarios may happen in the next time slot:
\begin{itemize}
\item
$i$ to $i+1$ ($0 \leq i \leq B_r-1$): the relay buffer is not full, and a packet arrives at the relay buffer.

\item
$i$ to $i-1$ ($1 \leq i \leq B_r$): the relay buffer is not empty, and a packet departures from the relay buffer.

\item
$i$ to $i$ ($0 \leq i \leq B_r$): no packet arrives at and departures from the relay buffer.
\end{itemize}

\begin{figure}[!t]
\centering
\includegraphics[width=3.0in]{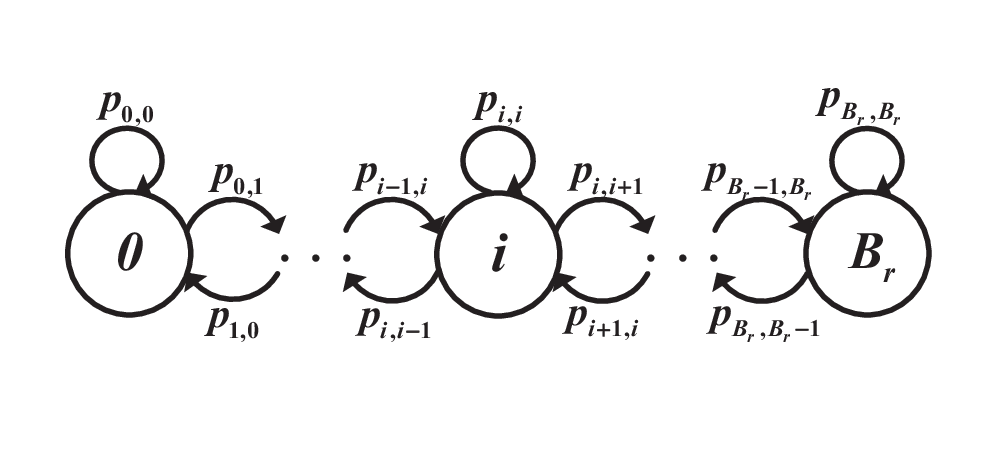} 
\caption{State machine of the birth-death chain.}
\label{fig:birth-death_chain}
\end{figure}

Let $p_{i,j}$ denote the one-step transition probability from state $i$ to state $j$ ($0 \leq i,j \leq B_r$), then the occupancy process $\{X_t,t=0,1,2,\cdots\}$ can be modeled as a birth-death chain as illustrated in Fig.~{\ref{fig:birth-death_chain}}. Let $\pi_r(i)$ denote the probability that there are $i$ packets occupying the relay buffer in the stationary state, the stationary OSD of the relay buffer $\mathbf{\Pi}_r=\left[\pi_r(0),\pi_r(1),\cdots,\pi_r(B_r) \right]$ is determined as 
\begin{align}
& \mathbf{\Pi}_r \mathbf{\cdot P}=\mathbf{\Pi}_r, \label{eq:balance_eq}  \\
& \mathbf{\Pi}_r \mathbf{\cdot 1} =1, \label{eq:normalization_eq}
\end{align}
where $\mathbf{P}$ is the one-step transition matrix of the birth-death chain defined as
\begin{equation}       
\mathbf{P}=\left[                 
\begin{array}{cccc}   
p_{0,0}   &   p_{0,1}   &                   &                                               \\  
p_{1,0}   &   p_{1,1}   & p_{1,2}           &                                               \\  
          &   \ddots    & \ddots            & \ddots                                        \\
          &             & p_{B_r,B_r-1}     & p_{B_r,B_r}
\end{array}
\right],    \label{eq:matrix}             
\end{equation}
and $\mathbf{1}$ is a column vector of size $(B_r+1)\times 1$ with all elements being 1. 

Notice that $p_{0,0}=1-p_{0,1}$, $p_{B_r,B_r}=1-p_{B_r,B_r-1}$ and $p_{i,i}=1-p_{i,i-1}-p_{i,i+1}$ for $0<i<B_r$, the expressions (\ref{eq:balance_eq})$-$(\ref{eq:matrix}) indicate that to derive $\mathbf{\Pi}_r$, we need to determine the one-step transition probabilities $p_{i,i+1}$ and $p_{i,i-1}$. 
\begin{lemma} \label{lemma:transition_probability}
For the birth-death chain in Fig.~\ref{fig:birth-death_chain}, its one-step transition probabilities $p_{i,i+1}$ and $p_{i,i-1}$ are determined as
\begin{align}
& p_{i,i+1}=p_{sr} \cdot (1-\pi_s(0)), \ 0 \leq i \leq B_r-1, \label{eq:p_i_i+1} \\
& p_{i,i-1}=p_{rd} \cdot \frac{i}{n-3+i}, \ 1 \leq i \leq B_r. \label{eq:p_i_i-1}
\end{align}
\end{lemma}

\begin{IEEEproof}
The proof is given in Appendix~\ref{appendix:transition_probability}.
\end{IEEEproof}

By substituting (\ref{eq:p_i_i+1}) and (\ref{eq:p_i_i-1}) into (\ref{eq:balance_eq}) and (\ref{eq:normalization_eq}), we can see that the stationary OSD of the relay buffer is determined as
\begin{align}
\pi_r(i)=\frac{\mathrm{C}_i (1-\pi_s(0))^i}{\sum\limits_{k=0}^{B_r}{\mathrm{C}_k (1-\pi_s(0))^k}}, \ 0 \leq i \leq B_r \label{eq:OSD_relay}
\end{align}
where $\mathrm{C}_i=\binom{n-3+i}{i}$.

\subsection{OSDs Analysis under the Scenario with Feedback}
Under the scenario with feedback, although node $\mathcal{S}$ gets the chance to execute the Source-to-Relay operation in a time slot, it still remains idle if the relay buffer of its intended receiver is full (with the overflow probability $\pi_r(B_r)$), which causes the correlation between the OSD analysis of source buffer and that of relay buffer. It is notable, however, the overflow probability $\pi_r(B_r)$ only affects the service rate $\mu_s$ of the source buffer and the arrival rate at the relay buffer, while the occupancy processes of the source buffer and relay buffer can still be modeled as the B/B/$1$/$B_s$ queue and the birth-death chain respectively. Thus, based on the similar analysis as that in Section~\ref{subsection:OSD_nofeedback}, we have the following corollary.
\begin{corollary} \label{corollary:OSD_feedback}
For the network scenario with feedback, the OSD $\mathbf{\Pi}_s$ of the source buffer and the OSD $\mathbf{\Pi}_r$ of the relay buffer are determined as (\ref{eq:OSD_source}) and ({\ref{eq:OSD_relay}}), where $\tau$ is given by (\ref{eq:tau}), and the service rate $\mu_s$ of the source buffer is evaluated as
\begin{equation}
\mu_s=p_{sd}+p_{sr} \cdot (1-\pi_r(B_r)). \label{eq:mu_s_FB}
\end{equation}
\end{corollary} 

\begin{IEEEproof}
The proof is given in Appendix~\ref{appendix:OSD_feedback}.
\end{IEEEproof}

Corollary~\ref{corollary:OSD_feedback} indicates that for the evaluation of OSDs $\mathbf{\Pi}_s$ and $\mathbf{\Pi}_r$, we need to determine the relay buffer overflow probability $\pi_r(B_r)$. From formula (\ref{eq:OSD_relay}) we have
\begin{equation}
\pi_r(B_r)=\frac{\mathrm{C}_{B_r} (1-\pi_s(0))^{B_r}} {\sum\limits_{k=0}^{B_r} \mathrm{C}_k (1-\pi_s(0))^k}, \label{eq:self-mapping}
\end{equation} 
where 
\begin{equation}
\pi_s(0)=\frac{\mu_s-\lambda_s^{\text{\scriptsize +}}} {\mu_s-\lambda_s^{\text{\scriptsize +}} \cdot \tau^{B_s}} = \frac{\mu_s-\lambda_s^{\text{\scriptsize +}}} {\mu_s-\lambda_s^{\text{\scriptsize +}} \cdot \left(\frac{\lambda_s^{\text{\scriptsize +}}(1-\mu_s)} {\mu_s (1-\lambda_s^{\text{\scriptsize +}})}\right)^{B_s}}. \label{eq:pi_s_0}
\end{equation}

We can see from (\ref{eq:mu_s_FB})$-$(\ref{eq:pi_s_0}) that (\ref{eq:self-mapping}) is actually an implicit function of $\pi_r(B_r)$, which can be solved by applying the fixed point theory \cite{Granas_BOOK03}. We provide in Appendix~\ref{appendix:fixed_point_iteration} the detailed fixed point iteration for solving $\pi_r(B_r)$.

\section{Performance Analysis} \label{section:performance}
With the help of OSDs of source buffer and relay buffer derived in Section~\ref{section:framework}, this section focuses on the performance analysis of the concerned buffer limited MANET in terms of its throughput, expected E2E delay and throughput capacity. 

\subsection{Throughput and Expected E2E Delay}

Regarding the throughput and expected E2E delay of a MANET with the general limited buffer constraint, we have the following theorem.

\begin{theorem} \label{theorem:throughput_delay}
For a concerned MANET with $n$ nodes, packet generating rate $\lambda_s^{\text{\scriptsize{+}}}$, source buffer size $B_s$ and relay buffer size $B_r$, its per flow throughput $T$ and expected E2E delay $\mathbb{E}\{D\}$ are given by
\begin{align}
& T =p_{sd}\left(1-\pi_s(0) \right)+p_{sr} \left(1-\pi_s(0) \right) (1-\pi_r(B_r)), \label{eq:throughput} \\
& \mathbb{E}\{D\}= \frac{1+L_s}{\mu_s}+\frac{(n-2+L_r)(1-\pi_r(B_r))}{p_{sd}+p_{sr}(1-\pi_r(B_r))}, \label{eq:E2E_delay}
\end{align}
where $L_s$ (resp. $L_r$) denotes the expected number of packets in the source buffer (resp. relay buffer) under the condition that the source buffer (resp. relay buffer) is not full, which is determined as
\begin{align}
& L_s=\frac{\tau-B_s\tau^{B_s}+(B_s-1)\tau^{B_s+1}}{(1-\tau)(1-\tau^{B_s})},  \label{eq:source_length}  \\ 
& L_r=\frac{1}{1-\pi_r(B_r)}\sum\limits_{i=0}^{B_r-1}{i\pi_r(i)}, \label{eq:relay_length}
\end{align}
$\mu_s$ is determined by (\ref{eq:mu_s_NF}) and (\ref{eq:mu_s_FB}) for the scenarios without and with feedback, respectively, $\tau$, $\pi_s(0)$ and $\mathbf{\Pi}_r$ are determined by (\ref{eq:tau}), (\ref{eq:OSD_source}) and (\ref{eq:OSD_relay}), respectively.
\end{theorem}

Notice that packets of a flow are delivered to their destination through either one-hop transmission (Source-to-Destination) or two-hop  transmission (Source-to-Relay and Relay-to-Destination), so the per flow throughput $T$ can be derived by analyzing packet delivery rates of these two kinds of transmissions. Regarding the expected E2E delay $\mathbb{E}\{D\}$, it can be evaluated based on the analysis of expected source queuing delay and expected delivery delay of a tagged packet\footnote{The source queuing delay of a packet is defined as the time it takes the packet to move to the HoL of its source queue after it is generated. The delivery delay of a packet is defined as the time it takes the packet to reach its destination after it moves to the HoL of its source queue.}. For the detailed proof of this theorem, please refer to Appendix~\ref{appendix:throughput_delay}.

\begin{remark}
The formulas (\ref{eq:throughput}) and (\ref{eq:E2E_delay}) hold for both network scenarios without/with feedback, but different network scenarios will lead to different results of $\tau$, $\pi_s(0)$ and $\mathbf{\Pi}_r$.
\end{remark}

Based on the results of Theorem~\ref{theorem:throughput_delay}, we can establish the following corollary (See Appendix~\ref{appendix:feedback} for the proof). 

\begin{corollary} \label{corollary:feedback}
For a concerned MANET with the general limited buffer constraint, adopting the feedback mechanism improves its throughput performance.
\end{corollary}

\subsection{Throughput Capacity and Limiting Throughput/Delay}
To determine the throughput capacity $T_c$, we first need the following lemma (See Appendix~\ref{appendix:as_lambda_increase} for the proof).

\begin{lemma} \label{lemma:as_lambda_increase}
For a concerned MANET with the general limited buffer constraint, its throughput $T$ increases monotonically as the packet generating rate $\lambda_s^{\text{\scriptsize +}}$ increases.
\end{lemma}

Based on Lemma~\ref{lemma:as_lambda_increase}, we can establish the following theorem on throughput capacity.

\begin{theorem} \label{theorem:throughput_capacity}
For a concerned MANET with $n$ nodes, source buffer size $B_s$ and relay buffer size $B_r$, its throughput capacity $T_c$ is given by 
\begin{equation}
T_c=p_{sd}+p_{sr}\frac{B_r}{n-2+B_r}. \label{eq:throughput_capacity}
\end{equation}
\end{theorem}

\begin{IEEEproof}
Lemma~\ref{lemma:as_lambda_increase} indicates that
\begin{equation}
T_c=\max\limits_{\lambda_s^{\text{\scriptsize{+}}}\in (0,1]} T = \lim\limits_{\lambda_s^{\text{\scriptsize +}} \to 1} T. \label{eq:Tc_lambda_1}
\end{equation}
From (\ref{eq:tau}), (\ref{eq:OSD_source}) and (\ref{eq:OSD_relay}) we can see that
\begin{equation}
\lim\limits_{\lambda_s^{\text{\scriptsize +}} \to 1} \tau=\lim\limits_{\lambda_s^{\text{\scriptsize +}} \to 1}\frac{\lambda_s^{\text{\scriptsize +}}(1-\mu_s)}{\mu_s(1-\lambda_s^{\text{\scriptsize +}})} \to \infty, 
\end{equation}
\begin{align}
\lim\limits_{\lambda_s^{\text{\scriptsize +}} \to 1} {\pi_s(0)} &=\lim\limits_{\lambda_s^{\text{\scriptsize +}} \to 1} \frac{\mu_s-\lambda_s^{\text{\scriptsize +}}}{\mu_s-\lambda_s^{\text{\scriptsize +}} \cdot \tau^{B_s}} \nonumber \\
&=\lim\limits_{\tau \to \infty}\frac{\mu_s-1}{\mu_s-\tau^{B_s}}=0, \label{eq:pi_s_0_lambda_1} 
\end{align}
\begin{align}
\lim\limits_{\lambda_s^{\text{\scriptsize +}} \to 1} {\pi_r(B_r)} & =\lim\limits_{\pi_s(0) \to 0} \frac{\mathrm{C}_{B_r} (1-\pi_s(0))^{B_r}}{\sum\limits_{k=0}^{B_r}{\mathrm{C}_k}(1-\pi_s(0))^k} \nonumber \\
& =\frac{\mathrm{C}_{B_r}}{\sum\limits_{k=0}^{B_r}{\mathrm{C}_k}}=\frac{n-2}{n-2+B_r}. \label{eq:pi_r_Br_lambda_1}
\end{align}
Then $T_c$ is given by 
\begin{align*}
T_c &  =\lim\limits_{\lambda_s^{\text{\scriptsize +}} \to 1} p_{sd}\left(1-\pi_s(0) \right)+p_{sr} \left(1-\pi_s(0) \right) (1-\pi_r(B_r)) \\
    &  =p_{sd} \cdot \left(1-0 \right)+p_{sr} \cdot \left(1-0 \right) \cdot \left(1-\frac{n-2}{n-2+B_r}\right)\\
		&  =p_{sd}+p_{sr}\frac{B_r}{n-2+B_r}.
\end{align*}
\end{IEEEproof}

\begin{remark}
We can see from Theorem~\ref{theorem:throughput_capacity} that the throughput capacity of the concerned MANET is the same for both the scenarios with and without feedback, and it is mainly determined by its relay buffer size $B_r$. We can further observe that when the network size $n$ is extremely large while the relay buffer size $B_r$ is fixed, the throughput is roughly equal to $p_{sd}$, which likes a unicast request only from a source node to its destination. This is because the service rate of a relay buffer is inversely proportional to the network size (please refer to Lemma~\ref{lemma:transition_probability}) and thus will tend to $0$ as $n$ increases, indicating that packets can hardly be forwarded to their destinations through a relay node.
\end{remark}

Based on Theorem~\ref{theorem:throughput_delay} and Theorem~\ref{theorem:throughput_capacity}, we have the following corollary regarding the limiting $T$ and $\mathbb{E}\{D\}$ as the buffer size tends to infinity (See Appendix~\ref{appendix:buffer_infinite} for the proof).
\begin{corollary} \label{corollary:buffer_infinite}
For a concerned MANET, its throughput increases as $B_s$ and/or $B_r$ increase. Moreover, as $B_s$ and/or $B_r$ tend to infinity, the corresponding limiting $T$ and $\mathbb{E}\{D\}$ are determined as (22) and (23) respectively (shown at the top of the next page), where $\rho_s=\min\{\frac{\lambda_s^{\text{\scriptsize +}}}{\mu_s}, 1\}$.
\begin{figure*}[!t]
\centering
\begin{subnumcases}{T=}
p_{sd} \cdot \rho_s + p_{sr} \cdot \frac{\sum\limits_{k=0}^{B_r-1} \mathrm{C}_k \rho_s^{k+1}}{\sum\limits_{k=0}^{B_r} \mathrm{C}_k \rho_s^k}, & $B_s \to \infty$ \label{eq:T_Bs_infinite}\\
(p_{sd}+p_{sr})(1-\pi_s(0)),  & $B_r \to \infty$  \label{eq:T_Br_infinite} \\
\min\{\lambda_s^{\text{\scriptsize +}},p_{sd}+p_{sr}\}. &  $B_s  \to  \infty$ and $B_r \to \infty$ \label{eq:T_Bs_Br_infinite}
\end{subnumcases} \label{eq:T_buffer_infinite}
\hrulefill
\end{figure*}
\begin{figure*}[!t]
\centering
\begin{subnumcases}{\mathbb{E}\{D\}=}
\infty, &$B_s \to \infty$ and $\lambda_s^{\text{\scriptsize +}} \geq \mu_s$ \\
\frac{1-\lambda_s^{\text{\scriptsize +}}}{\mu_s-\lambda_s^{\text{\scriptsize +}}} + \frac{(n-2+L_r)(1-\pi_r(B_r))}{p_{sd}+p_{sr}(1-\pi_r(B_r))}, &
	   $B_s \to \infty$ and $\lambda_s^{\text{\scriptsize +}} < \mu_s$   \label{eq:D_Bs_infinite}   \\
\frac{n-2+\pi_s(0) \cdot (1+L_s)}{\pi_s(0) \cdot (p_{sd}+p_{sr})},  & $B_r \to \infty$ \label{eq:D_Br_infinite} \\
\frac{n-1-\lambda_s^{\text{\scriptsize +}}}{p_{sd}+p_{sr}-\lambda_s^{\text{\scriptsize +}}}, & $B_s \to \infty$, $B_r \to \infty$ and $\lambda_s^{\text{\scriptsize +}}<\mu_s$	 \label{eq:D_Bs_Br_infinite}
\end{subnumcases}
\hrulefill
\end{figure*}
\end{corollary}

\begin{remark}
Corollary~\ref{corollary:buffer_infinite} indicates the throughput and delay results derived in Theorem 1 are universal in the sense that they cover the concise forms derived in other works as special cases. For example, they reduce to the results in \cite{Liu_ADHOC15,Liu_TWC15} as $B_s$ tends to infinity, and the results in \cite{Neely_IT05} as both $B_s$ and $B_r$ tend to infinity.
\end{remark}

\section{Case Studies} \label{section:case_studies}
In the previous two sections, with the basic probabilities $p_{sd}$, $p_{sr}$ and $p_{rd}$, we have developed a theoretical framework for the performance modeling of a general MANET. These probabilities are determined by the specific MAC protocol adopted. To demonstrate the application and efficiency of our framework, in this section, we conduct case studies under network scenarios with two typical MAC protocols widely used in other studies concerning MANETs, and present corresponding theoretical/simulation results.

\subsection{Network Scenarios}

\textbf{Cell-partitioned MANET with Local Scheduling based MAC (LS-MAC) \cite{Neely_IT05,Sharma_TON07,Wang_TON11,Urgaonkar_TON11}:}
Under this network scenario, the whole network area is evenly partitioned into $m \times m$ non-overlapping cells. In each time slot one cell supports only one transmission between two nodes within it, and concurrent transmissions in different cells will not interfere with each other. When there are more than one node in a cell, each node in this cell becomes the transmitter equally likely. For such a  MANET, the corresponding probabilities $p_{sd}$, $p_{sr}$ and $p_{rd}$ can be determined by the following formulas (See Appendix~\ref{appendix:basic_probabilities} for derivations).
\begin{equation}
p_{sd}=\frac{m^2}{n}-\frac{m^2-1}{n-1}+\frac{m^2-1}{n(n-1)}\left(1-\frac{1}{m^2}\right)^{n-1}, \label{eq:p_sd_LS}
\end{equation} 
\begin{align}
p_{sr} &= p_{rd} \nonumber \\
       &= \frac{1}{2}\left\{\frac{m^2\!-\!1}{n\!-\!1}\!-\!\frac{m^2}{n\!-\!1}\left(1\!-\!\frac{1}{m^2}\right)^n\!-\!\left(1\!-\!\frac{1}{m^2}\right)^{n\!-\!1} \right\}.  \label{eq:p_sr_LS}
\end{align}

\begin{figure}
    \centering        
    \subfigure[Transmission range of a node.]	
		{\includegraphics[width=1.65in]{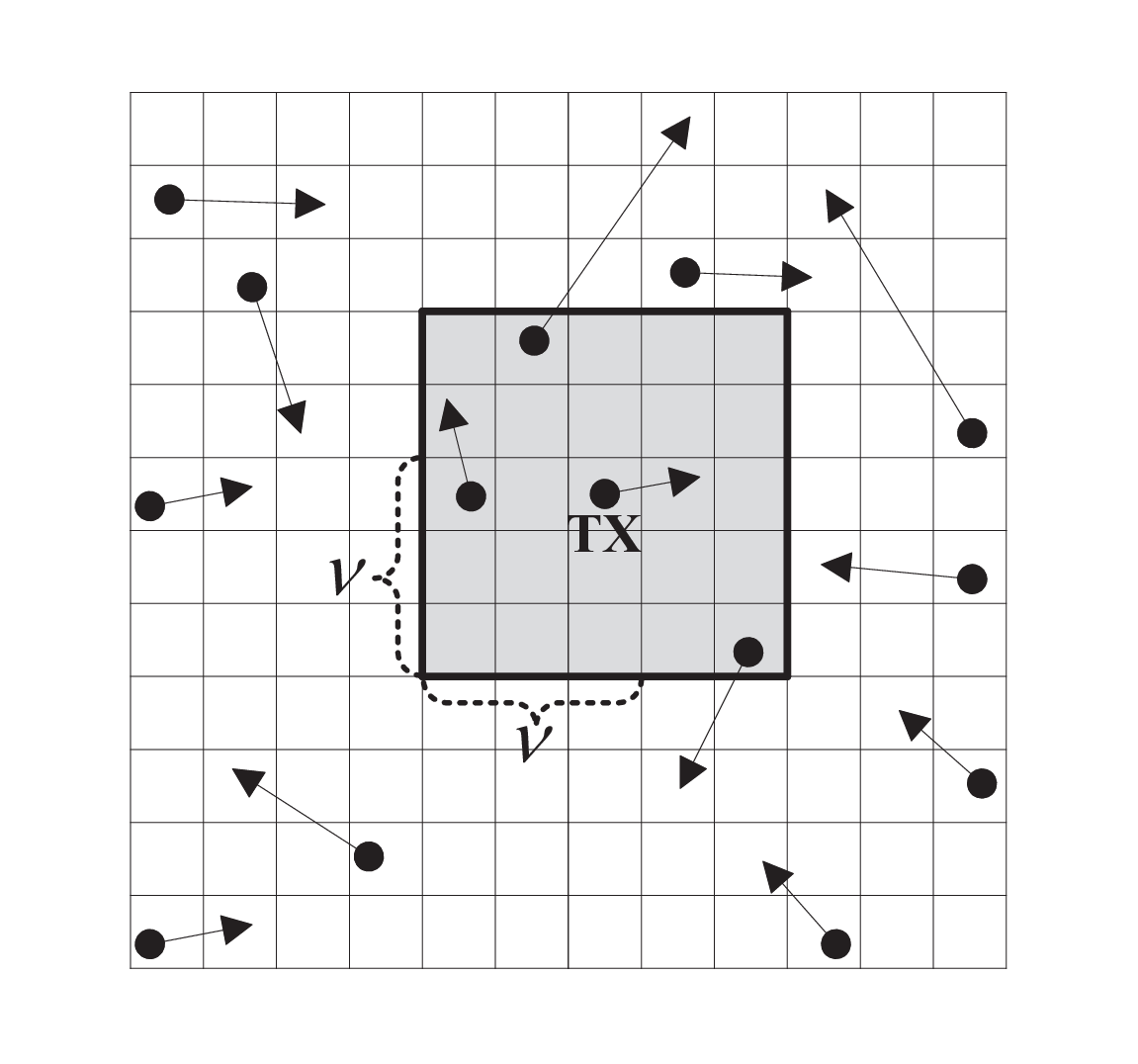} \label{fig:transmission_range} }
		\hfill
    \subfigure[Illustration of an EC (all the cells with gray color belong to the same EC).]		
		{\includegraphics[width=1.65in]{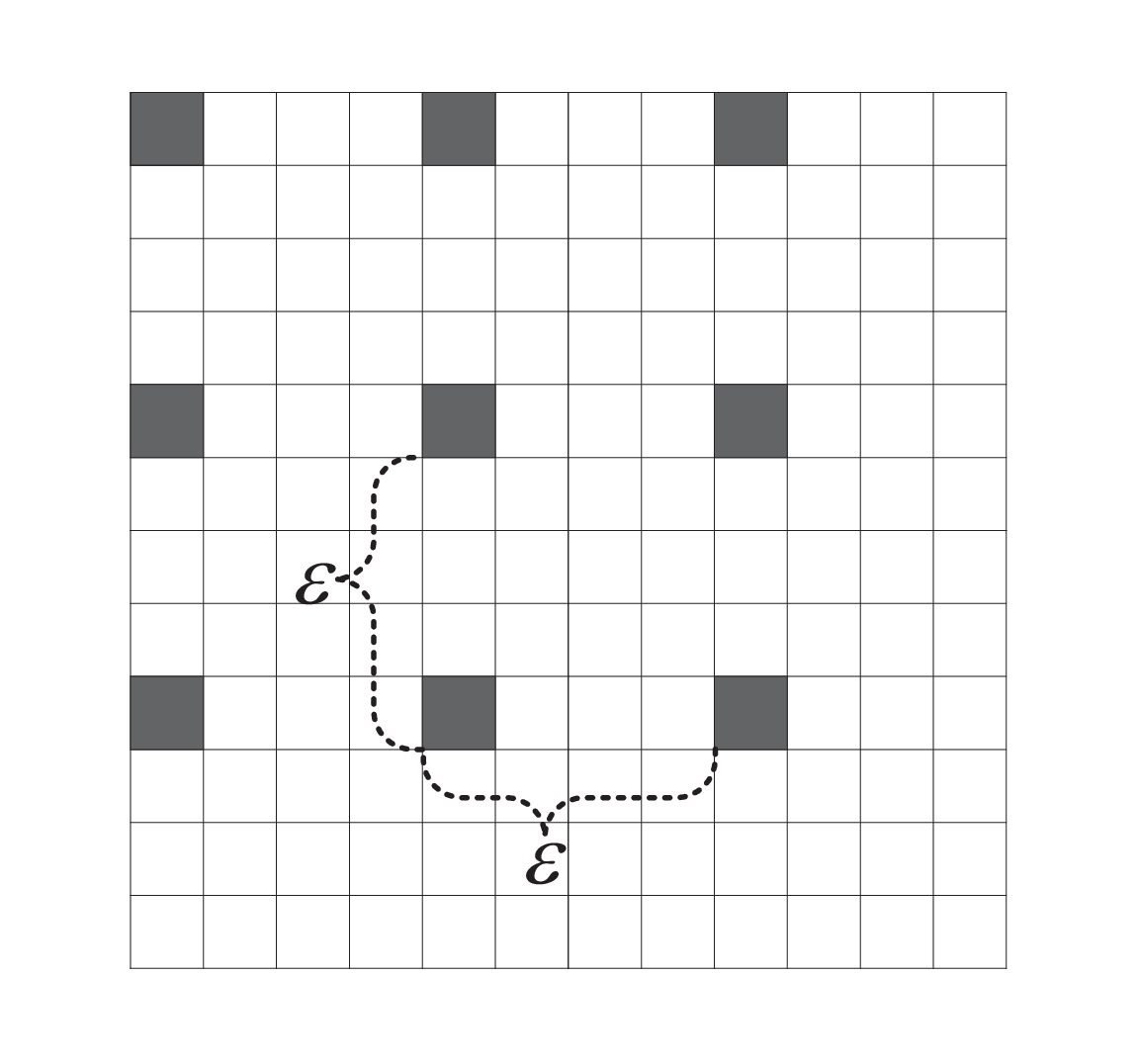}\label{fig:equivalence_class}}
    \caption{A cell-partitioned MANET with EC-MAC.}
\end{figure}

\textbf{Cell-partitioned MANET with Equivalence Class based MAC (EC-MAC) \cite{Herdtner_INFOCOM05,Kulkarni_IT04,Franceschetti_IT07,Shakkottai_TON10}:}
In such a MANET, the whole network area is evenly partitioned into $m \times m$ non-overlapping cells, and each transmitter (like the $TX$ in Fig.~\ref{fig:transmission_range}) has a transmission range that covers a set of cells with horizontal and vertical distance of no more than $\nu-1$ cells away from the cell the transmitter resides in. To prevent simultaneous transmissions from interfering with each other, the EC-MAC is adopted. As illustrated in Fig.~\ref{fig:equivalence_class} that with the EC-MAC, all cells are divided into different ECs, and any two cells in the same EC have a horizontal and vertical distance of some multiple of $\varepsilon$ cells. Each EC alternatively becomes active every $\varepsilon^2$ time slots, and each active cell of an active EC allows only one node in it (if any) to conduct data transmission. When there are more than one node in an active cell, each node in this cell becomes the transmitter equally likely. To enable as many number of concurrent transmissions to be scheduled as possible while avoiding interference among these transmissions, $\varepsilon$ should be set as \cite{Liu_TWC15}
\begin{equation}
\varepsilon=\min \{\lceil (1+\Delta) \sqrt{2} \nu+\nu \rceil,m \}, \label{eq:epsilon}
\end{equation}
where $\Delta$ is a guard factor specified by the protocol model \cite{Gupta_IT00}.

For such a MANET, the corresponding probabilities $p_{sd}$, $p_{sr}$ and $p_{rd}$ are determined by the following formulas (See Appendix~\ref{appendix:basic_probabilities} for derivations).
\begin{equation}
p_{sd}\!= \!\frac{1}{\varepsilon^2} \left \{ \frac{\Gamma\!-\!\frac{m^2}{n}}{n\!-\!1} \!+\! \frac{m^2\!-\!1\!-\!(\Gamma\!-\!1)n}{n(n\!-\!1)} \left(1\!-\!\frac{1}{m^2}\right) ^{n\!-\!1} \right \},  \label{eq:p_sd_EC} \\
\end{equation}
\begin{align}
p_{sr} &\!=\!p_{rd} \nonumber \\
       &\!=\!\frac{1}{2 \varepsilon^2} \left \{ \frac{m^2\!-\!\Gamma}{n\!-\!1} \left(1\!-\!\left(1\!-\!\frac{1}{m^2}\right)^{n\!-\!1}\right) \!- \!\left(1\!-\!\frac{\Gamma}{m^2}\right)^{n\!-\!1}     \right \}, \label{eq:p_sr_EC}
\end{align}
where $\Gamma=(2\nu-1)^2$.

By substituting the results of (\ref{eq:p_sd_LS})-(\ref{eq:p_sr_LS}) and (\ref{eq:epsilon})-(\ref{eq:p_sr_EC}) into our theoretical framework, the network performance of a cell-partitioned MANET with LS-MAC and EC-MAC can be obtained, respectively. Our framework can easily apply to any other MAC protocol. For example, $p_{sd}$, $p_{sr}$ and $p_{rd}$ were derived for MANETs with Aloha protocol in \cite{Chen_TCOM16}, then the performance modeling of Aloha MANETs under the general limited buffer constraint can be accordingly conducted.

\subsection{Simulation Settings}

\begin{table*}[t]
\centering
\caption{Simulation Settings}
\begin{tabular}{|c|c|c|c|c|c|c|c|c|c|}
\hline
Parameters & $n$  & $m$ & $B_s$ & $B_r$ & $\nu$ & $\Delta$ & time slots      & media access control & mobility model                 \tabularnewline \hline
Settings   & $72$ & $6$ & $5$   & $5$   & $1$   & $1$      & $2 \times 10^8$ & LS-MAC and EC-MAC    & i.i.d model and RW model    \tabularnewline \hline
\end{tabular}
\label{tab:settings}
\end{table*}

To validate our theoretical framework for MANET performance modeling, a simulator was developed to simulate the packet generating, packet queuing and packet delivery processes under above two network scenarios \cite{C++}. Each simulation task runs over a period of $2 \times 10^8$ time slots, and we only collect data from the last $80 \%$ of time slots to ensure the system is in the steady state. In the simulator, the following two typical mobility models have been implemented:
\begin{itemize}
\item 
\textbf{I.i.d Model \cite{Neely_IT05}:}
At the beginning of each time slot, each node independently selects a cell among all cells with equal probability and then stays in it during this time slot.
 
\item
\textbf{Random Walk (RW) Model \cite{Gamal_IT06}:}
At the beginning of each time slot, each node independently selects a cell among its current cell and its $8$ adjacent cells with equal probability $1/9$ and then stays in it during this time slot.
\end{itemize}

The detailed settings of network parameters in our simulations are summarized in Table~\ref{tab:settings}. Readers can also flexibly perform our C++ simulator with any other desired parameter settings.

\subsection{Theoretical/Simulation Results} \label{subsection:validation}
\begin{figure}[!t]
    \centering
    {
    \subfigure[LS-MAC: $T$ versus $\lambda_s^{\text{\scriptsize +}}$.]
    {\includegraphics[width=1.65in]{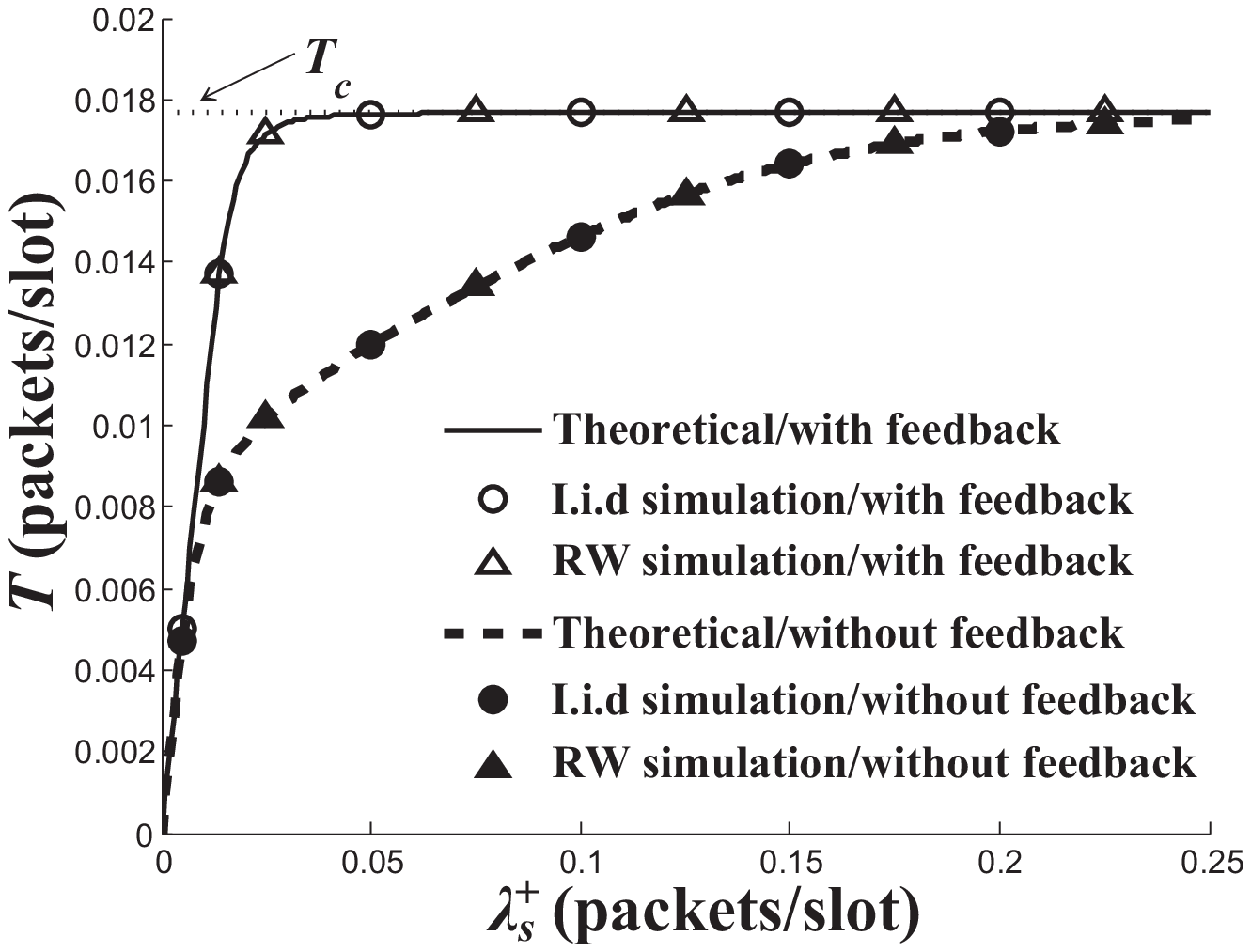} \label{fig:throughput_LS}}
   	\hfill
    \subfigure[LS-MAC: $\mathbb{E}\{D\}$ versus $\lambda_s^{\text{\scriptsize +}}$.]
    {\includegraphics[width=1.65in]{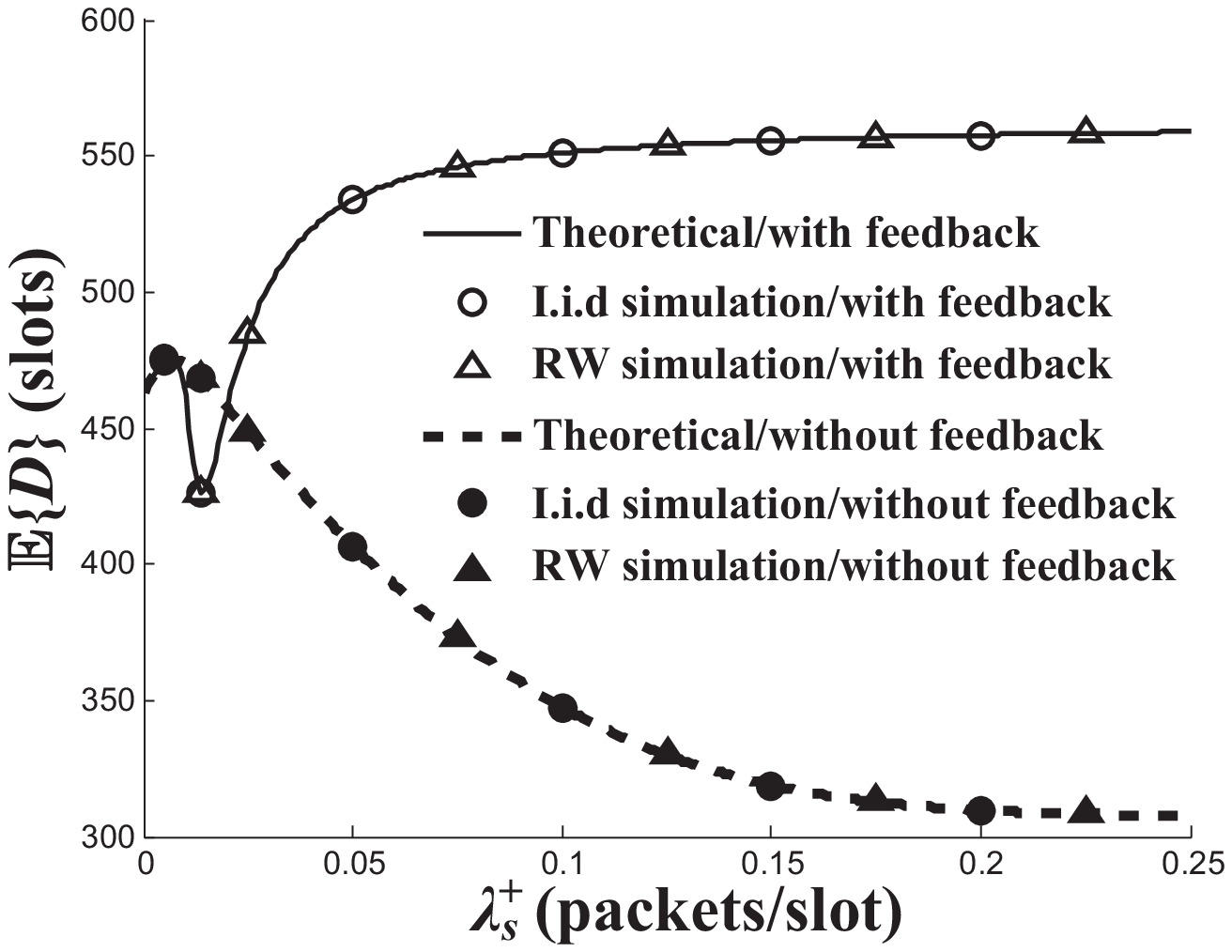} \label{fig:delay_LS}}
		\hfill
		\subfigure[EC-MAC: $T$ versus $\lambda_s^{\text{\scriptsize +}}$.]
    {\includegraphics[width=1.65in]{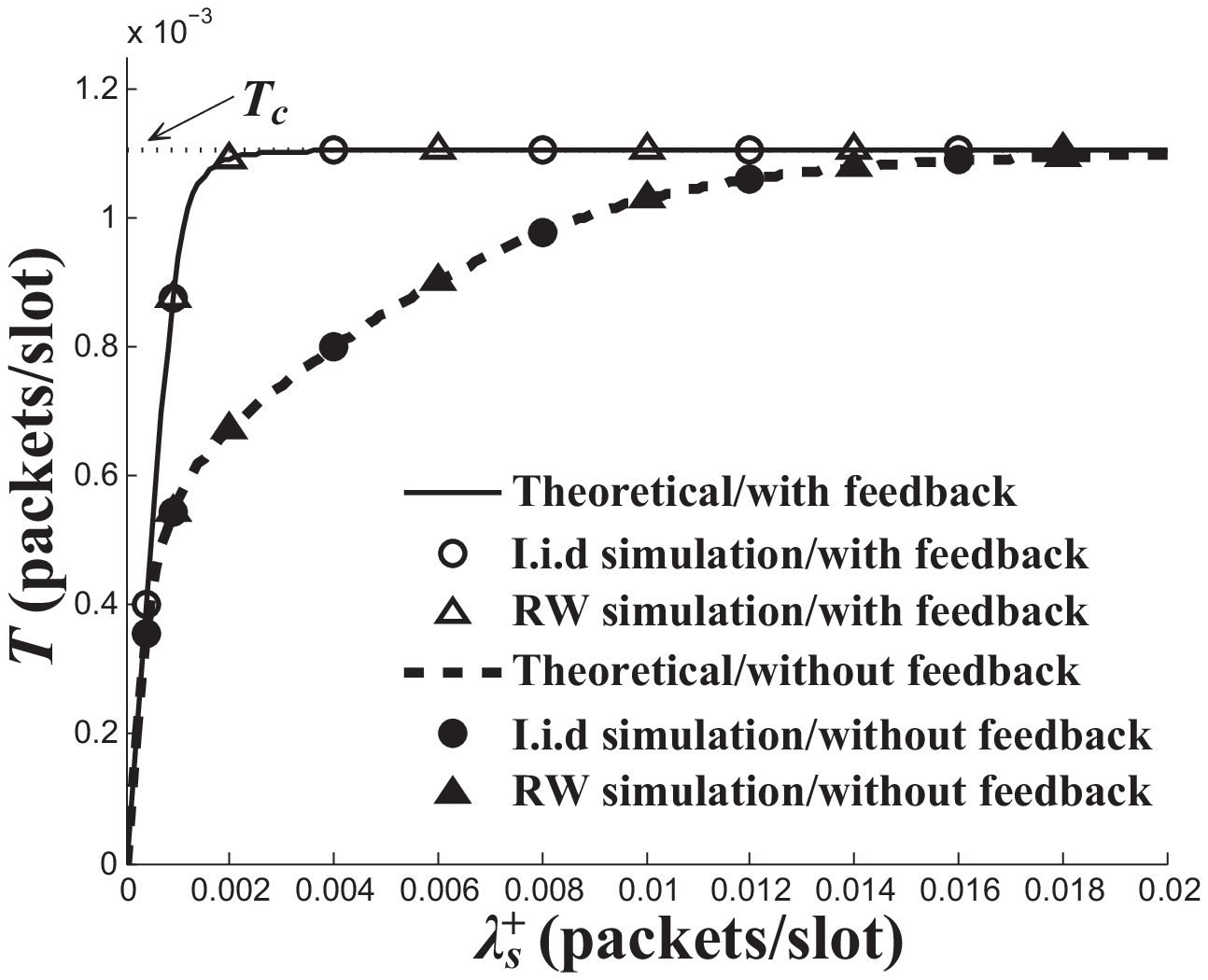} \label{fig:throughput_EC}}
   	\hfill
    \subfigure[EC-MAC: $\mathbb{E}\{D\}$ versus $\lambda_s^{\text{\scriptsize +}}$.]
    {\includegraphics[width=1.65in]{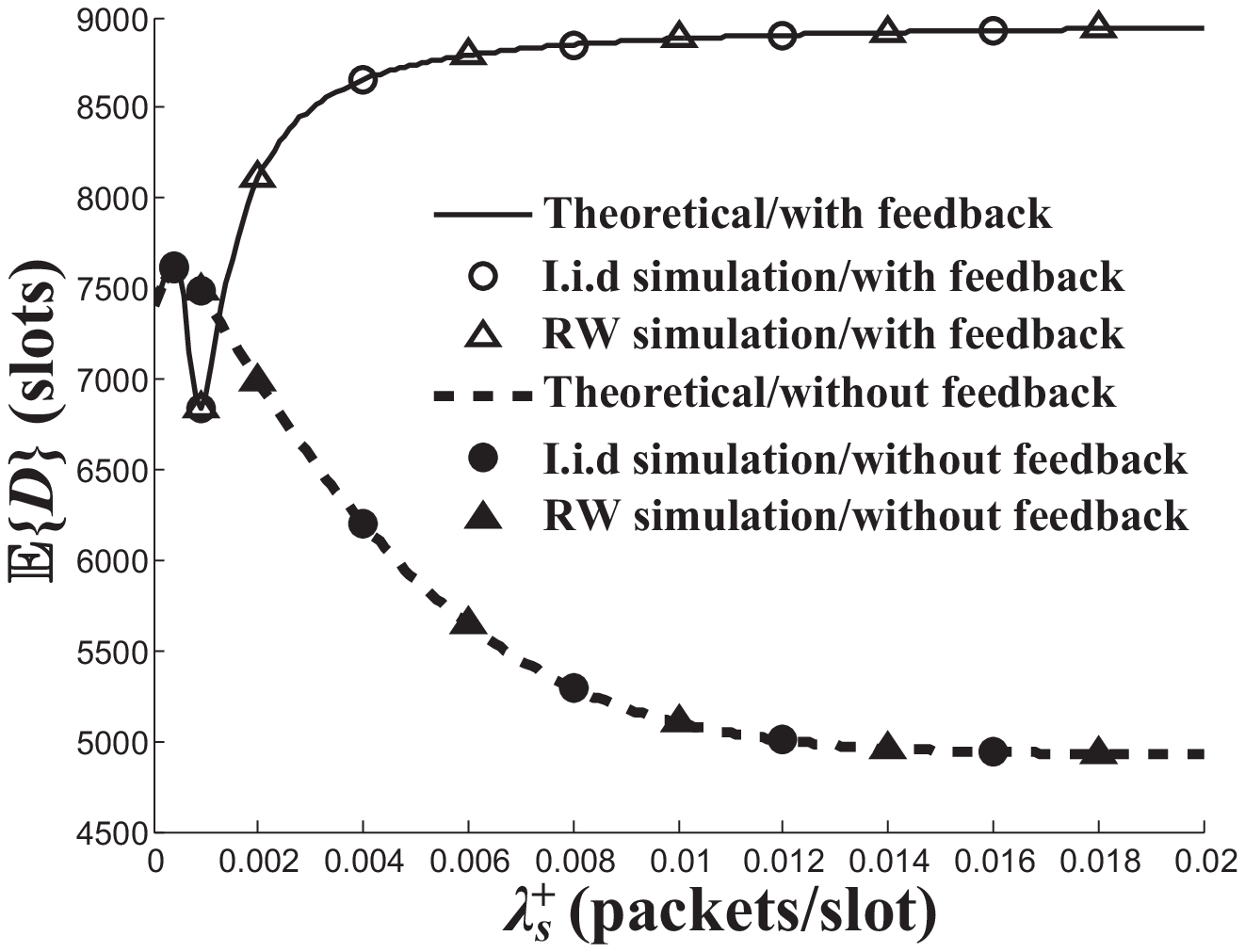} \label{fig:delay_EC}}
    }
    \caption{Performance validation.}
    \label{fig:validation}
\end{figure}

We summarize in Fig.~\ref{fig:validation} the theoretical/simulation results for throughput and delay under the above two network scenarios. For each scenario we consider the network settings of ($n=72,m=6,B_s=5,B_r=5$), and for the scenario with the EC-MAC protocol we set $\nu=1$ and $\Delta=1$ there \cite{ns-2}. Notice that the theoretical results here are obtained by substituting (\ref{eq:p_sd_LS}) and (\ref{eq:p_sr_LS}) (resp. (\ref{eq:p_sd_EC}) and (\ref{eq:p_sr_EC})) into the formulas in Theorem~\ref{theorem:throughput_delay}.

Fig.~\ref{fig:validation} shows clearly that the simulation results match well with the theoretical ones for all the cases considered here, which indicates that our theoretical framework is applicable to and highly efficient for the performance modeling of different buffer limited MANETs. We can see from Fig.~\ref{fig:throughput_LS} and Fig.~\ref{fig:throughput_EC} that for a MANET with LS-MAC or EC-MAC, as the packet generating rate $\lambda_s^{\text{\scriptsize +}}$ increases, the per flow throughput $T$ increases monotonically and finally converges to its throughput capacity $T_c$, which agrees with the conclusions of Lemma~\ref{lemma:as_lambda_increase} and Theorem~\ref{theorem:throughput_capacity}. Another interesting observation of Fig.~\ref{fig:throughput_LS} and Fig.~\ref{fig:throughput_EC} is that just as predicated by  Corollary~\ref{corollary:feedback} and Theorem~\ref{theorem:throughput_capacity}, although adopting the feedback mechanism usually leads to a higher throughput, it does not improve the throughput capacity performance.

Regarding the delay performance, we can see from Fig.~\ref{fig:delay_LS} and Fig.~\ref{fig:delay_EC} that in a MANET with either LS-MAC or EC-MAC, the behavior of expected E2E delay $\mathbb{E}\{D\}$ under the scenario without feedback is quite different from that under the scenario with feedback. As $\lambda_s^{\text{\scriptsize +}}$ increases, in the scenario without feedback $\mathbb{E}\{D\}$ first slightly increases and then decreases monotonically, while in the scenario with feedback $\mathbb{E}\{D\}$ first slightly increases, then decreases somewhat and finally increases monotonically. This is due to the reason that $\mathbb{E}\{D\}$ consists of source queuing delay and delivery delay, and the effects of $\lambda_s^{\text{\scriptsize +}}$ on $\mathbb{E}\{D\}$ are two folds. On one hand, a larger $\lambda_s^{\text{\scriptsize +}}$ leads to a more congested network with a larger $\pi_r(B_r)$ and a smaller $\mu_s$ (see formula~(\ref{eq:mu_s_FB})), which further leads to a larger expected source queuing delay; on the other hand, a larger $\pi_r(B_r)$ indicates that a packet is more likely to be delivered through a direct Source-to-Destination transmission, which further leads to a smaller expected delivery delay. As $\lambda_s^{\text{\scriptsize +}}$ increases, either of the two effects becomes dominant alternatively, causing the increase-decrease-increase phenomena of $\mathbb{E}\{D\}$ (it can be also seen in Fig.~\ref{fig:D_Bs_lambda} and Fig.~\ref{fig:D_Br_lambda} later).

Moreover, the results in Fig.~\ref{fig:validation} indicate that although adopting the feedback mechanism leads to an improvement in per flow throughput, such improvement usually comes with a cost of a larger E2E delay. This is because that the feedback mechanism can avoid the packet dropping at a relay node, which contributes to the throughput improvement but at the same time makes the source/relay buffers tend to be more congested, leading to an increase in delay.


\section{Numerical Results and Discussions} \label{section:numerical_results}

Based on the proposed theoretical framework, this section presents extensive numerical results to illustrate the potential impacts of buffer constraint on network performance. Notice from Section~\ref{subsection:validation} that the performance behaviors of the LS-MAC are quite similar to that of the EC-MAC, in the following discussions we only focus on a MANET with the LS-MAC.

\begin{figure}
    \centering        
    \subfigure[$T$ versus $B_s$.]	
		{\includegraphics[width=1.65in]{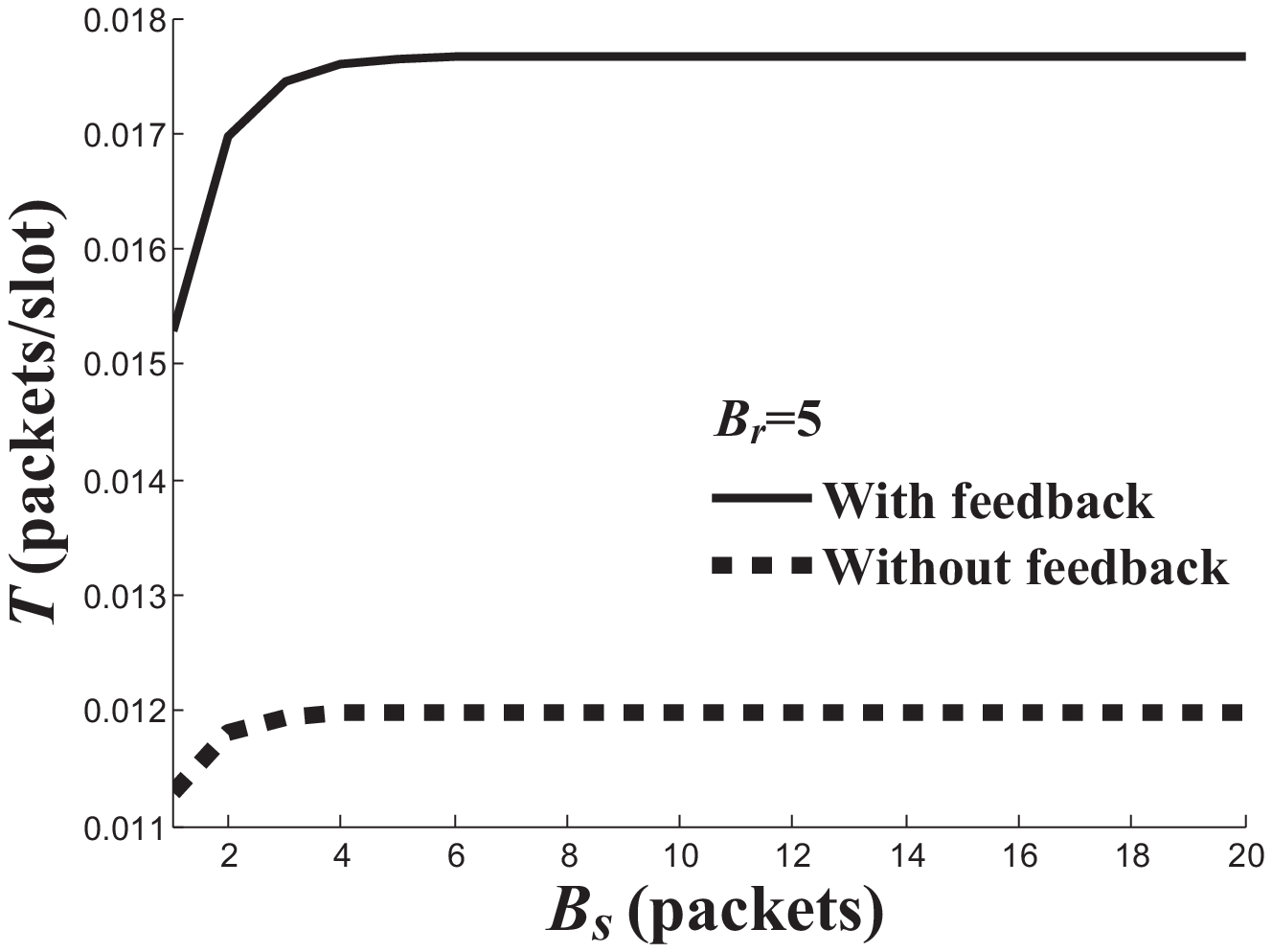} \label{fig:throughput_vs_Bs} }
		\hfill
    \subfigure[$\mathbb{E}\{D\}$ versus $B_s$.]		
		{\includegraphics[width=1.65in]{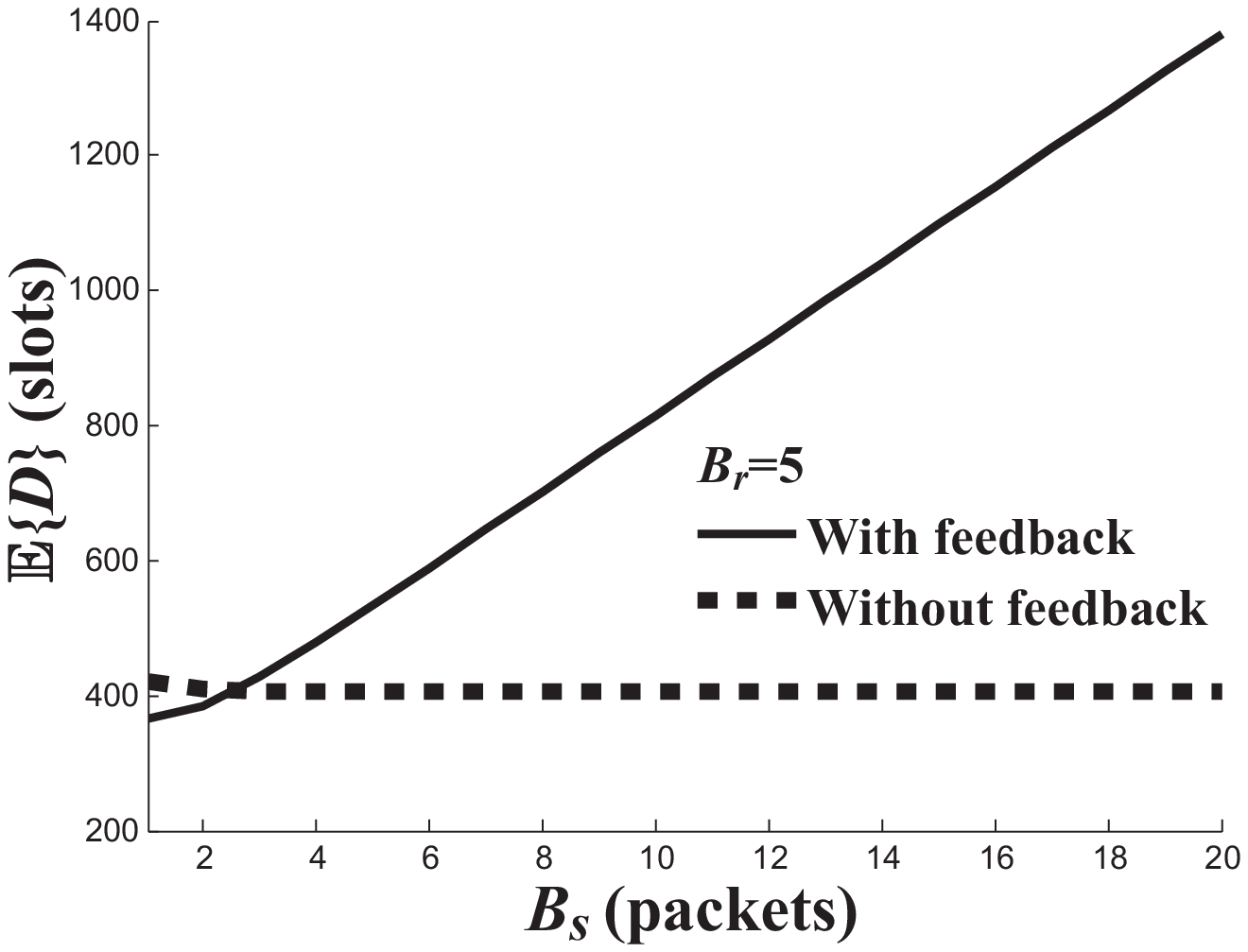} \label{fig:delay_vs_Bs}}
		\hfill
    \subfigure[$T$ versus $B_r$.]		
		{\includegraphics[width=1.65in]{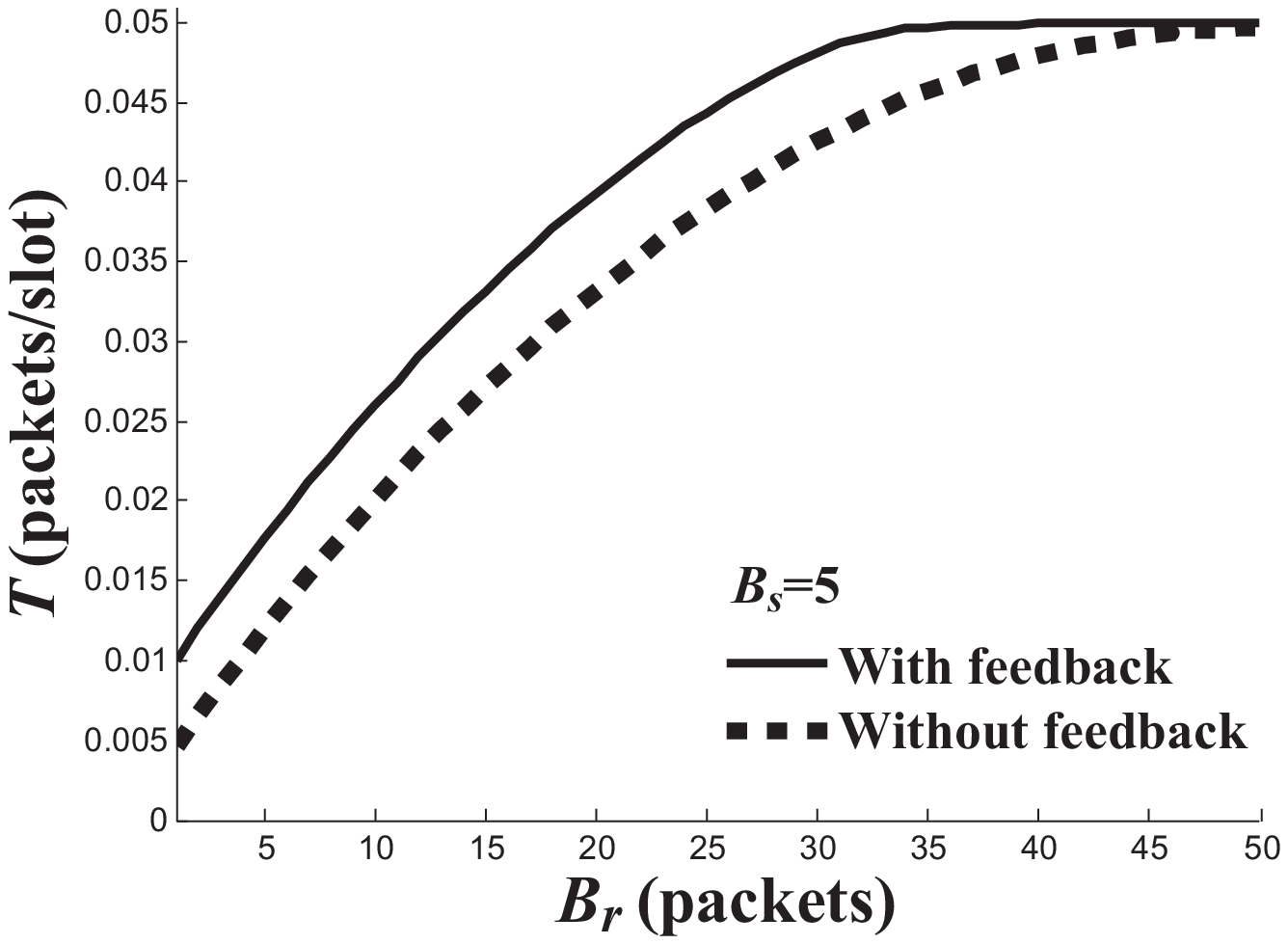} \label{fig:throughput_vs_Br}}
		\hfill
    \subfigure[$\mathbb{E}\{D\}$ versus $B_r$.]		
		{\includegraphics[width=1.65in]{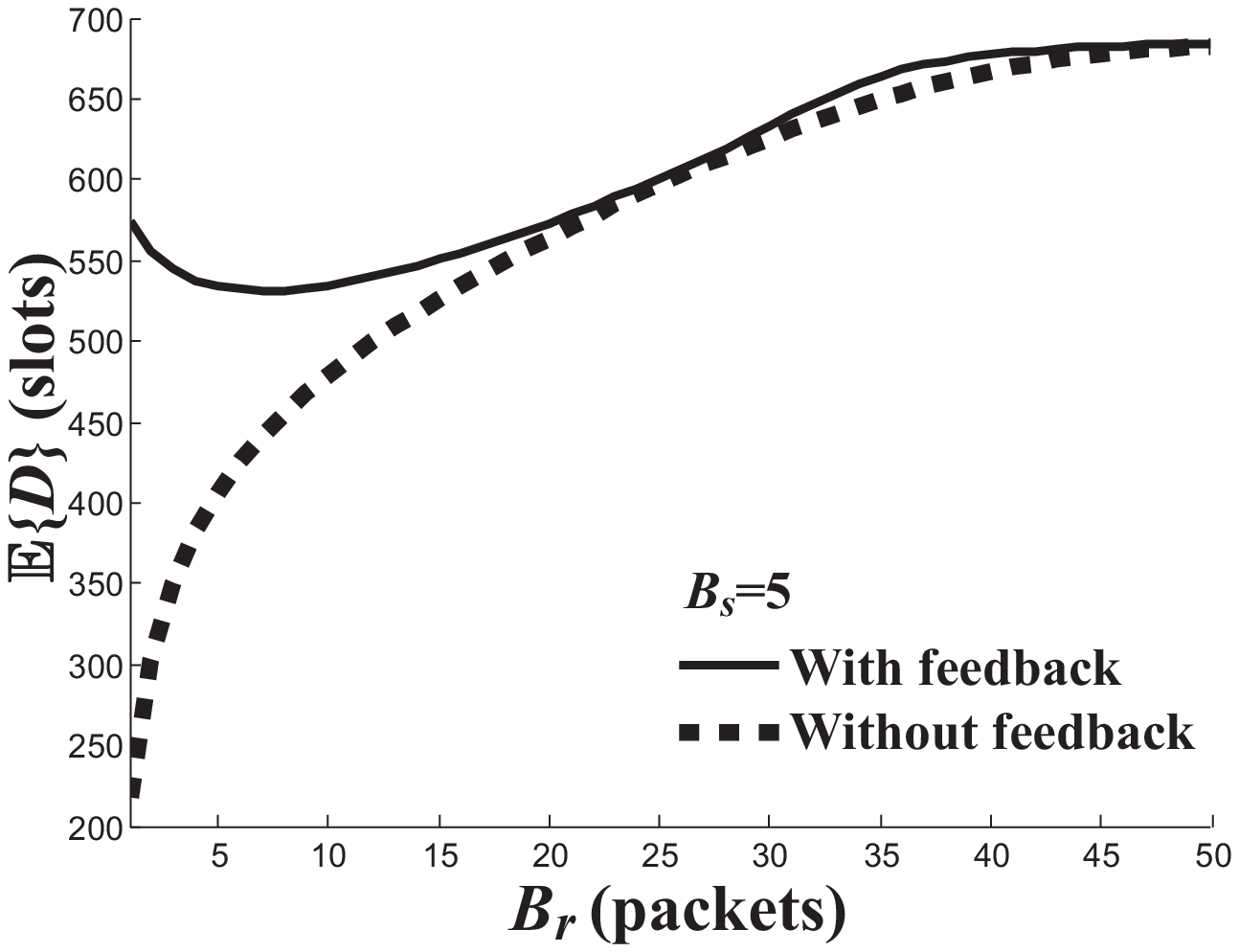} \label{fig:delay_vs_Br}}
		\caption{Throughput and delay versus $B_s$ and $B_r$ for the network setting of ($n=72$, $m=6$, $\lambda_s^{\text{\scriptsize +}}=0.05$).}
		\label{fig:T_D_vs_Bs_Br}
\end{figure}

We first summarize in Fig.~\ref{fig:T_D_vs_Bs_Br} how $T$ and $\mathbb{E}\{D\}$ vary with $B_s$ and $B_r$ under the settings of ($n=72$, $m=6$, $\lambda_s^{\text{\scriptsize +}}=0.05$). About the throughput performance, we can see from Fig.\ref{fig:throughput_vs_Bs} and Fig.\ref{fig:throughput_vs_Br} that just as predicated by Corollary~\ref{corollary:buffer_infinite} and Corollary~\ref{corollary:feedback}, $T$ increases as either $B_s$ or $B_r$ increases, and the feedback mechanism can lead to an improvement in $T$. It is interesting to see that as $B_s$ increases, $T$ under the two scenarios without and with feedback converges to two distinct constants determined by (22a). As $B_r$ increases, however, $T$ under the two scenarios finally converges to the same constant determined by (22b). 

Regarding the delay performance, Fig.~\ref{fig:delay_vs_Bs} shows that as $B_s$ increases, $\mathbb{E}\{D\}$ under the scenario without feedback quickly converges to a constant determined by (23b), while $\mathbb{E}\{D\}$ under the scenario with feedback monotonically increases to infinity, which agrees with the result of (23a). We can see from Fig.~\ref{fig:delay_vs_Br} that with the increase of $B_r$, however, $\mathbb{E}\{D\}$ under the scenario without feedback monotonically increases, while $\mathbb{E}\{D\}$ under the scenario with feedback first decreases and then increases. This is due to the reason that the effects of $B_r$ on $\mathbb{E}\{D\}$ are also two folds. On one hand, a larger $B_r$ leads to a less congested network with a smaller $\pi_r(B_r)$ and a larger $\mu_s$, which further leads to a smaller expected source queuing delay; on the other hand, a larger $B_r$ indicates that a packet is more likely to be delivered through a two-hop way (Source-to-Relay and Relay-to-Destination), which leads to a larger expected delivery delay. Similar to the throughput behavior in Fig.~\ref{fig:throughput_vs_Br}, Fig.~\ref{fig:delay_vs_Br} shows that as $B_r$ increases $\mathbb{E}\{D\}$ under the two scenarios also converges to the same constant determined by (23c).

The results in Fig.~\ref{fig:T_D_vs_Bs_Br} indicate that $B_s$ and $B_r$ have different impacts on the network performance in terms of $T$ and $\mathbb{E}\{D\}$. In particular, as $B_s$ increases, a notable performance gap between the scenarios without and with feedback always exist, where the throughput gap converges to a constant but the corresponding delay gap tends to infinity. As $B_r$ increases, however, the performance gap between the two scenarios tends to decrease to $0$, which implies that the benefits of adopting the feedback mechanism are diminishing in MANETs with a large relay buffer size.  
A further careful observation of Fig.~\ref{fig:T_D_vs_Bs_Br} indicates that although we can improve the throughput by increasing $B_s$ or $B_r$, it is more efficient to adopt a large $B_r$ rather than a large $B_s$  for such improvement. For example, under the scenario without feedback, Fig.~\ref{fig:throughput_vs_Bs} shows that by increasing $B_s$ from $1$ to $20$, $T$ can be improved from $0.0113$ to $0.0120$ (with an improvement of $6.19\%$); while Fig.~\ref{fig:throughput_vs_Br} shows that by increasing $B_r$ from $1$ to $20$, $T$ can be improved from $0.0046$ to $0.0332$ (with an improvement of $621.74\%$). 

\begin{figure}
    \centering        
    \subfigure[$T$ versus ($\lambda_s^{\text{\scriptsize +}},B_s$), $B_r=5$.]	
		{\includegraphics[width=1.65in]{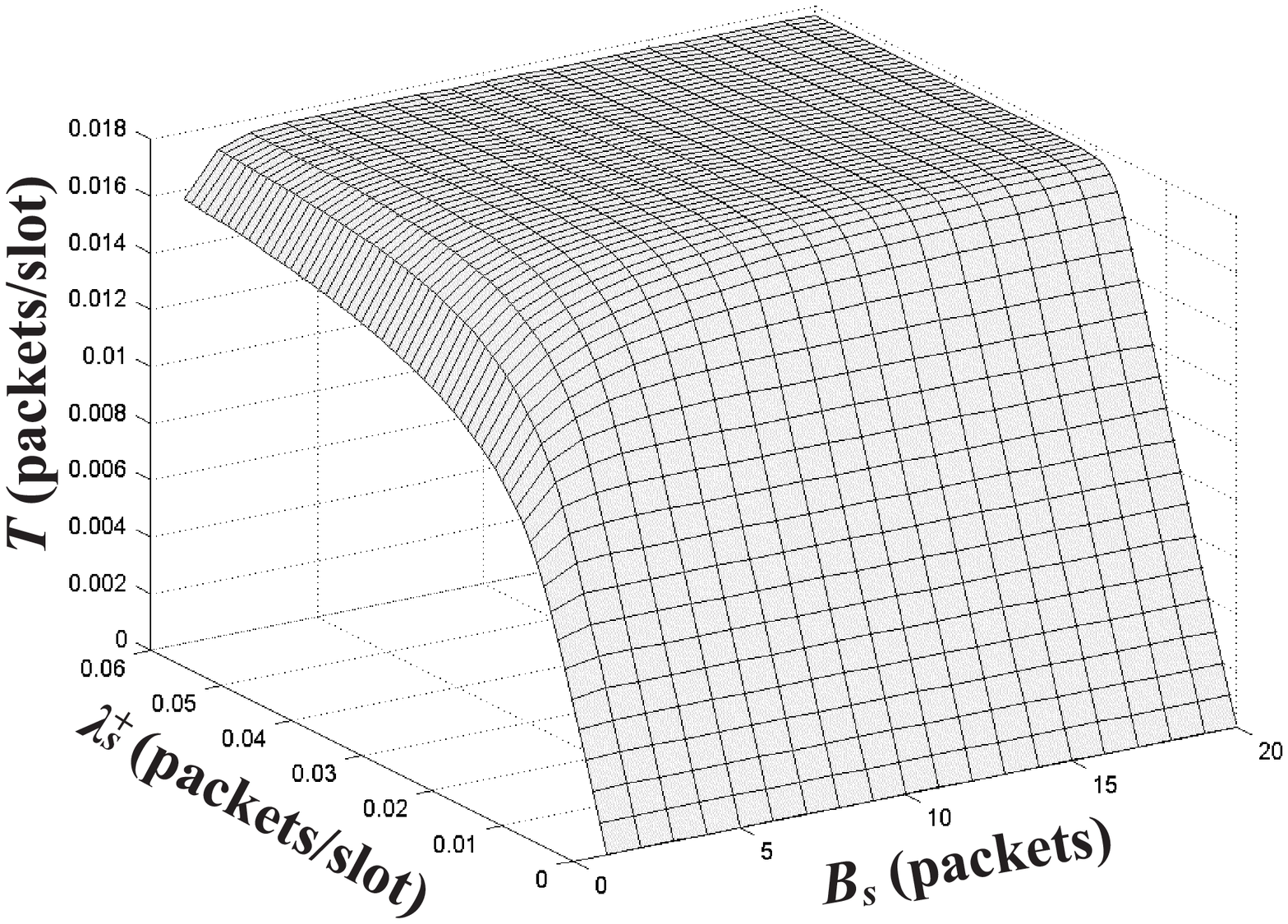} \label{fig:T_Bs_lambda} }
		\hfill
		\subfigure[$\mathbb{E}\{D\}$ versus ($\lambda_s^{\text{\scriptsize +}},B_s$), $B_r=5$.]	
		{\includegraphics[width=1.65in]{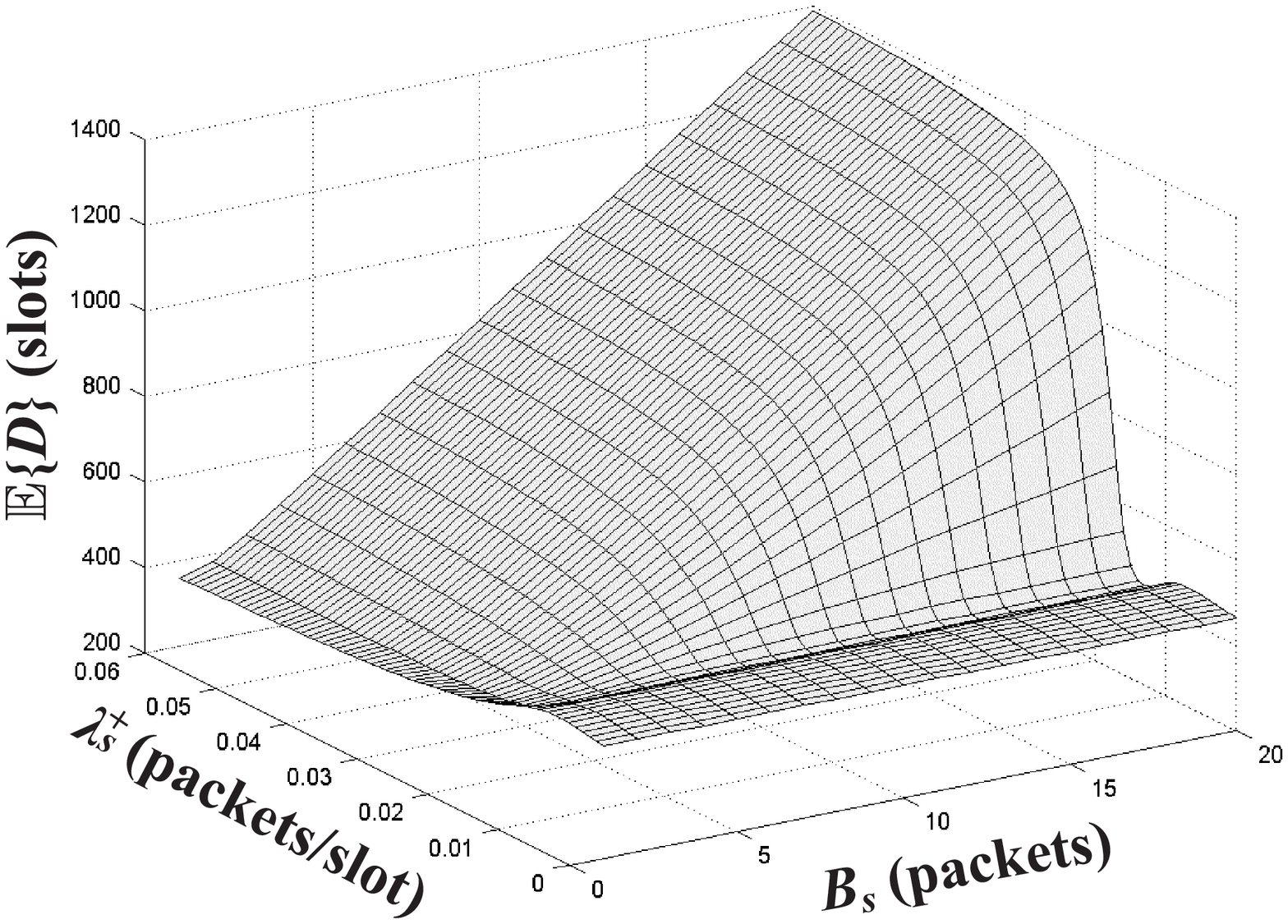} \label{fig:D_Bs_lambda} }
		\hfill
		\subfigure[$T$ versus ($\lambda_s^{\text{\scriptsize +}},B_r$), $B_s=5$.]	
		{\includegraphics[width=1.65in]{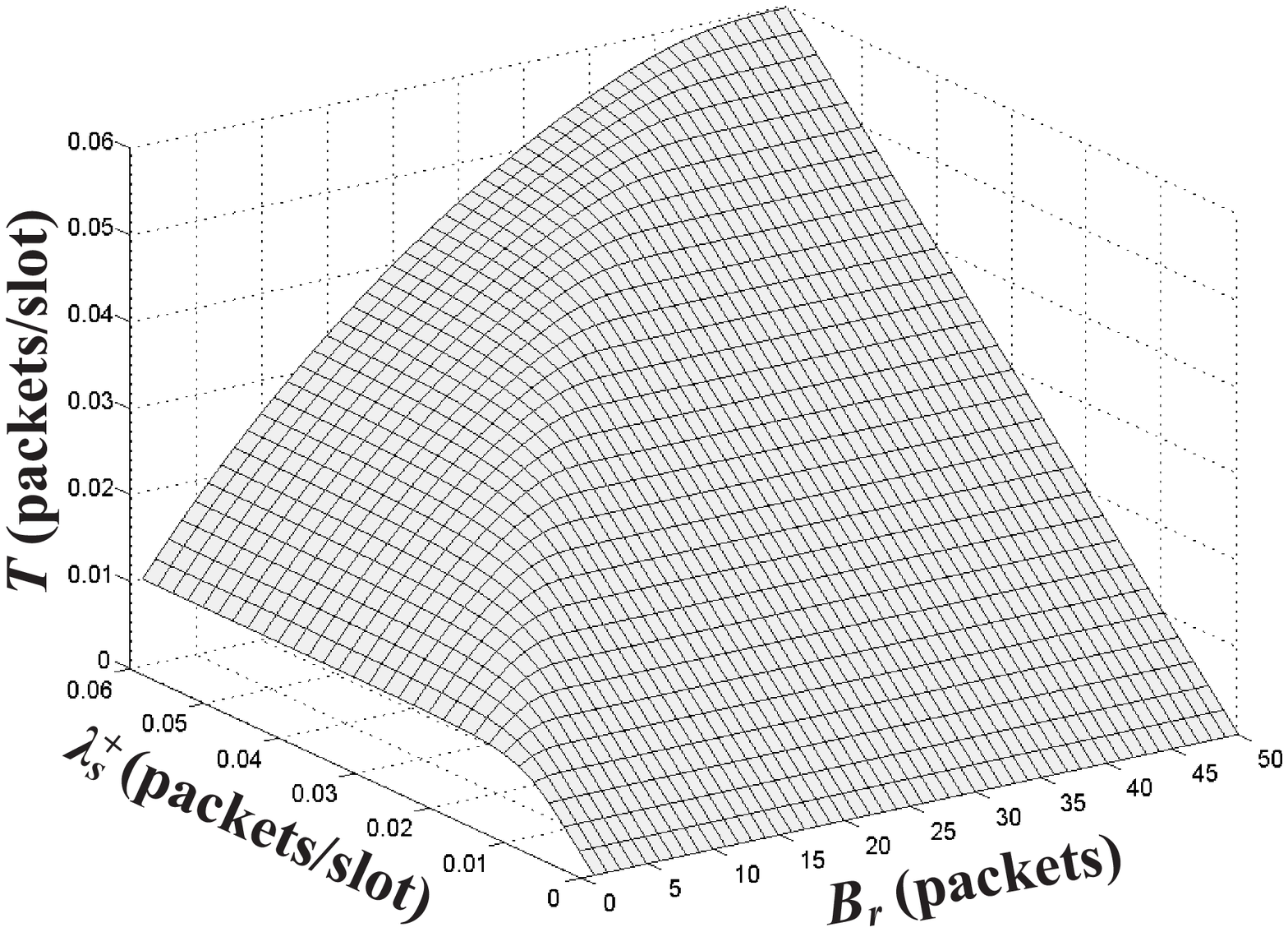} \label{fig:T_Br_lambda} }
		\hfill
		\subfigure[$\mathbb{E}\{D\}$ versus ($\lambda_s^{\text{\scriptsize +}},B_r$), $B_s=5$.]	
		{\includegraphics[width=1.65in]{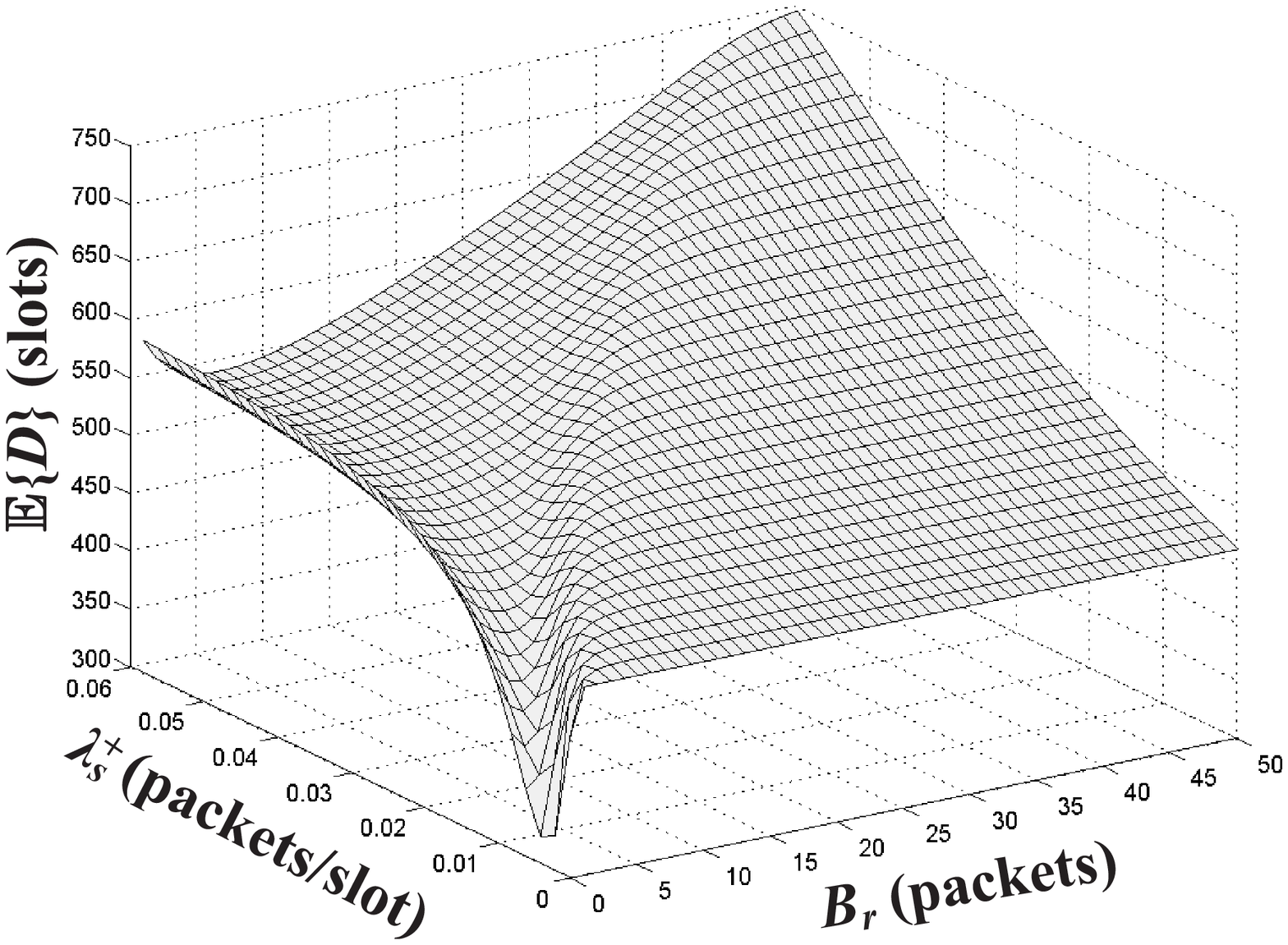} \label{fig:D_Br_lambda} }
		\caption{Throughput and delay versus ($\lambda_s^{\text{\scriptsize +}},B_s$) and ($\lambda_s^{\text{\scriptsize +}},B_r$) for the network setting of ($n=72$, $m=6$).}
		\label{fig:T_D_vs_Bs_Br_lambda}
\end{figure}

To further illustrate how the impacts of buffer size on network performance are dependent on packet generating rate $\lambda_s^{\text{\scriptsize +}}$, we focus on a MANET with feedback and summarize in Fig.~\ref{fig:T_D_vs_Bs_Br_lambda} how its throughput and delay vary with $\lambda_s^{\text{\scriptsize +}}$ and ($B_s,B_r$). We can see from Fig.~\ref{fig:T_Bs_lambda} and Fig.~\ref{fig:T_Br_lambda} that although in general we can improve $T$ by increasing either $B_s$ or $B_r$, the degree of such improvement is highly dependent on $\lambda_s^{\text{\scriptsize +}}$. As $\lambda_s^{\text{\scriptsize +}}$ increases, the throughput improvement from $B_r$ monotonically increases, while the corresponding improvement from $B_s$ first increases and then decreases. Fig.~\ref{fig:T_Bs_lambda} and Fig.~\ref{fig:T_Br_lambda} also show that as $\lambda_s^{\text{\scriptsize +}}$ increases, $T$ under different settings of $B_s$ finally converges to the same constant (i.e., $T_c$ given by (\ref{eq:throughput_capacity})), while $T$ under a given setting of $B_r$ converges to a distinct constant of $T_c$, which monotonically increases as $B_r$ increases.

Regarding the joint impacts of $\lambda_s^{\text{\scriptsize +}}$ and $B_s$ on delay performance, we can see clearly from Fig.~\ref{fig:D_Bs_lambda} that just as discussed in Corollary~\ref{corollary:buffer_infinite}, there exists a threshold of $\lambda_s^{\text{\scriptsize +}}$ beyond which $\mathbb{E}\{D\}$ will increases to infinity as $B_s$ increases, while for a given $\lambda_s^{\text{\scriptsize +}}$ less than the threshold, $\mathbb{E}\{D\}$ almost keeps as a constant as $B_s$ increases. About the joint impacts of $\lambda_s^{\text{\scriptsize +}}$ and $B_r$ on delay performance, Fig.~\ref{fig:D_Br_lambda} shows that for a given setting of $\lambda_s^{\text{\scriptsize +}}$, there also exists a threshold for $B_r$, beyond which $\mathbb{E}\{D\}$ almost keeps as a constant as $B_r$ increases. It is interesting to see that such threshold for $B_r$ and the corresponding delay constant tend to increase as $\lambda_s^{\text{\scriptsize +}}$ increases. The results in Fig.~\ref{fig:D_Br_lambda} imply that a bounded $\mathbb{E}\{D\}$ can be always guaranteed in a MANET as long as its source buffer size is limited.

\begin{figure}
    \centering        
    \subfigure[$T$ versus $n$.]	
		{\includegraphics[width=0.23\textwidth]{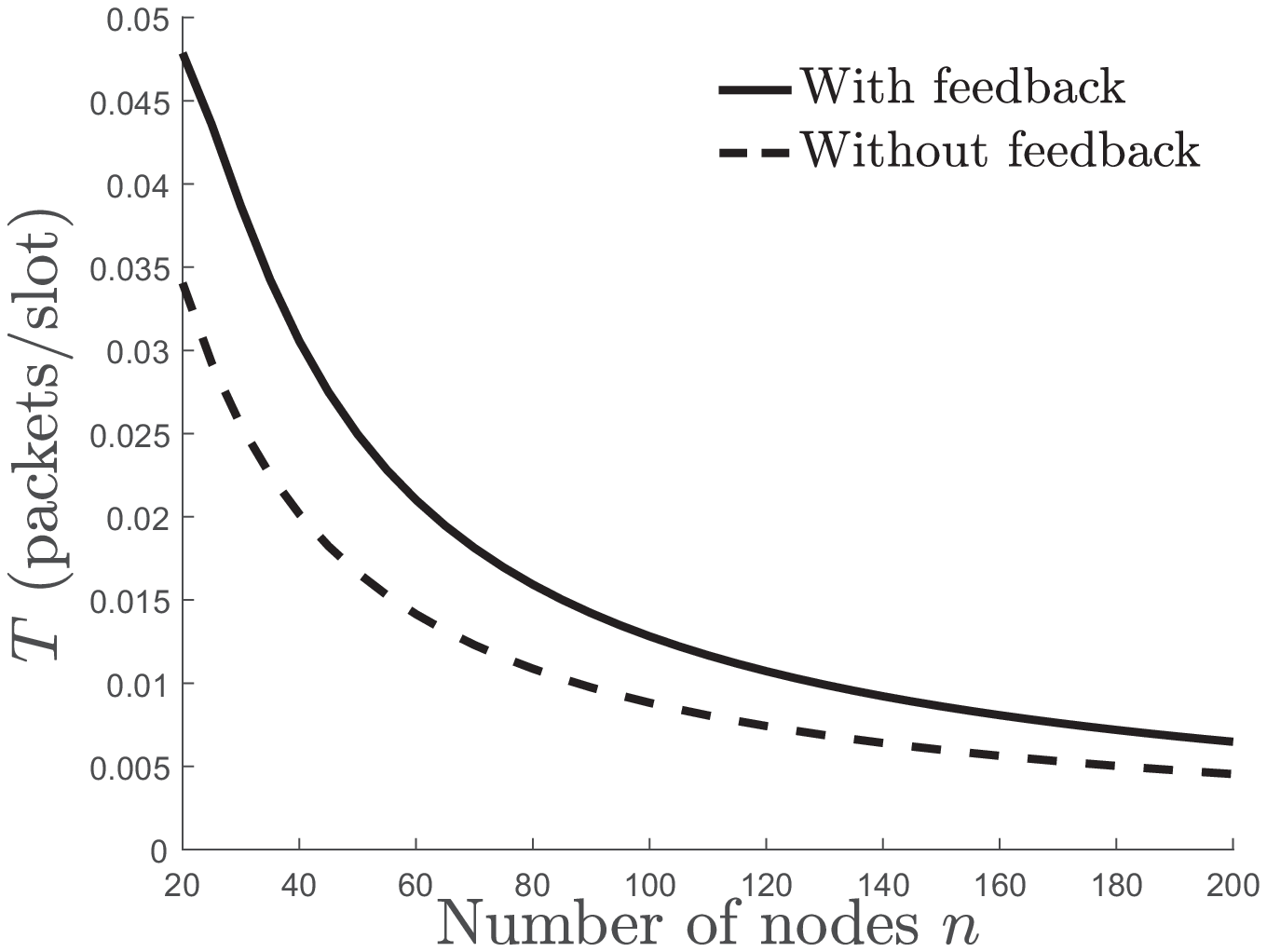} \label{fig:T_n} }
		\hfill
		\subfigure[$\mathbb{E}\{D\}$ versus $n$.]	
		{\includegraphics[width=0.23\textwidth]{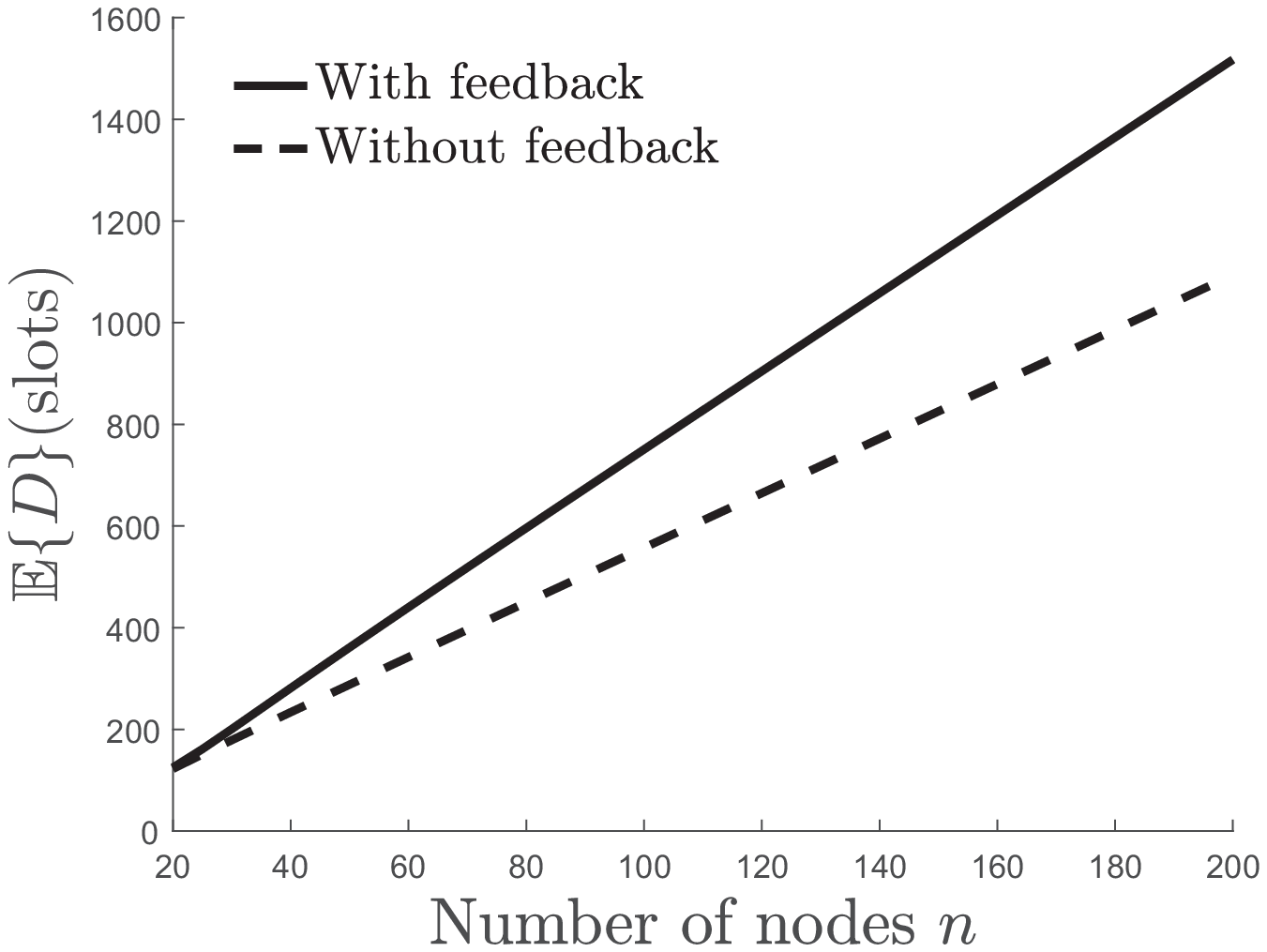} \label{fig:D_n} }
		\caption{Throughput and delay versus the number of nodes $n$ for the network setting of ($B_s=5$, $B_r=5$, $\lambda_s^{\text{\scriptsize +}}=0.05$, $d=2$).}
		\label{fig:T_D_vs_n}
\end{figure}

Finally, we plot Fig.~\ref{fig:T_D_vs_n} to illustrate the network performance behaviors as the number of nodes $n$ increases, where we set $B_s=5$, $B_r=5$, $\lambda_s^{\text{\scriptsize +}}=0.05$ and $d=2$ ($d$ denotes the node/cell density). We can see from Fig.~\ref{fig:T_n} that for both the network scenarios without and with feedback, the per flow throughput $T$ decreases monotonically as $n$ increases. When $n$ tends to infinity, from (\ref{eq:p_sd_LS}) and (\ref{eq:p_sr_LS}) we have $p_{sd}$ and $p_{sr}$ tend to $0$ and $\frac{1-e^{-d}-de^{-d}}{2d}$, respectively, and from (\ref{eq:throughput_capacity}) we can further observe that the throughput capacity scales as $\Theta(B_r/n)$. It indicates that to achieve a non-vanishing per flow throughput in a
MANET under general limited buffer constraint, the relay buffer size $B_r$ should grow at least linearly with the number of nodes $n$. Regarding the delay performance, Fig.~\ref{fig:D_n} shows that for both the network scenarios without and with feedback, $\mathbb{E}\{D\}$ increases almost linearly with $n$. This linear growth behavior can be also observed in other works such as \cite{Neely_IT05,Gamal_IT06,Sharma_TON07}, while the new insight revealed here is that the cost of adopting feedback mechanism to improve throughput performance is a steeper growth slope of E2E delay with the network size. 

\section{Conclusion} \label{section:conclusion}

This paper explored, for the first time, the performance modeling for MANETs under the general limited buffer constraint. In particular, a complete and generally applicable theoretical framework was developed to capture the inherent buffer occupancy behaviors in such a MANET, which enables the exact expressions to be derived for some fundamental network performance metrics, like the achievable throughput, expected E2E delay and throughput capacity. Some interesting conclusions that can be drawn from this study are: 1) In general, adopting the feedback mechanism can lead to an improvement in the throughput performance, but such improvement comes with the cost of a relatively large delay; 2) For the purpose of throughput improvement, it is more efficient to adopt a large relay buffer rather than a large source buffer; 3) The throughput capacity is dominated by the relay buffer size rather than the source buffer size; 4) Feedback mechanism cannot improve the throughput capacity.

Notice that in this paper, only buffer constraint was investigated, so one promising future direction is to conduct performance study for MANETs under more practical network scenarios, where the packet loss could be caused by other reasons such as poor signal conditions. Another appealing future direction is to explore the performance modeling for MANETs with the retransmission scheme.

\appendices

\section{Proof of Lemma~\ref{lemma:transition_probability}} \label{appendix:transition_probability}

Based on the transition scenarios, we can see $p_{i,i+1}$ is actually equal to the packet arrival rate $\lambda_r^{\text{\scriptsize +}}$ of the relay buffer, so we just need to determine $\lambda_r^{\text{\scriptsize +}}$ for the evaluation of $p_{i,i+1}$. When $\mathcal{S}$ serves as a relay, all other $n-2$ nodes (except $\mathcal{S}$ and its destination $\mathcal{D}$) may forward packets to it. When one of these nodes sends out a packet from its source buffer, it will forward the packet to $\mathcal{S}$ with probability $\frac{p_{sr}}{\mu_s(n-2)}$. This is because with probability $\frac{p_{sr}}{\mu_s}$ the packet is intended for a relay node, and each of the $n-2$ relay nodes are equally likely. Thus,
\begin{equation}
p_{i,i+1}=\lambda_r^{\text{\scriptsize +}}=(n-2) \lambda_s^{\text{\small -}} \cdot \frac{p_{sr}}{\mu_s (n-2)}, \label{eq:lambda_r+}
\end{equation}
where $\lambda_s^{\text{\small -}}$ denotes the packet departure rate of a source buffer. Due to the reversibility of the B/B/1/{\small$B_s$} queue, the packet departure process of the source buffer is also a Bernoulli process with its departure rate $\lambda_s^{\text{\small -}}$ being determined as
\begin{equation}
\lambda_s^{\text{\small -}}=\mu_s\left(1-\pi_s(0) \right). \label{eq:lambda_s-}
\end{equation} 
Then we have
\begin{equation*}
p_{i,i+1}=\lambda_r^{\text{\scriptsize +}}=p_{sr} \cdot (1-\pi_s(0)), \ 0 \leq i \leq B_r-1. 
\end{equation*}

\begin{figure}[!t]
\centering
\includegraphics[width=3.0in]{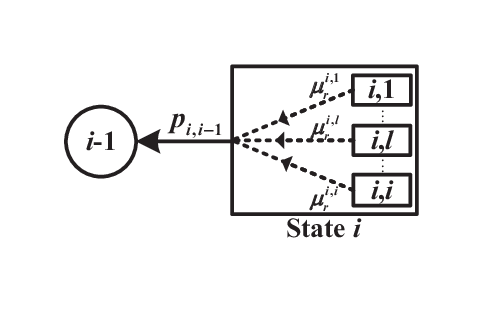} 
\caption{Illustration of the state decomposition.}
\label{fig:state_breakdown}
\end{figure}

Regarding the evaluation of transition probability $p_{i,i-1}$, it is notable that $p_{i,i-1}$ just corresponds to the service rate $\mu_r^i$ of the relay buffer when it is at state $i$. To determine $\mu_r^i$, we further decompose the state $i$ ($i>0$) into $i$ sub-states $\{(i,l), 1 \leq l \leq i\}$ as illustrated in Fig.~\ref{fig:state_breakdown}, where $l$ denotes the number of non-empty relay queues in the relay buffer. Let $\mu_r^{i,l}$ denote the service rate of the relay buffer when it is at sub-state $(i,l)$, and let $P_{l|i}$ denote the probability that the relay buffer is at sub-state $(i,l)$ conditioned on that the relay buffer is at state $i$, we then have 
\begin{equation}
\mu_r^i=\sum\limits_{l=1}^i {P_{l|i}\cdot \mu_r^{i,l}}. \label{eq:mu_r_i}
\end{equation}

We first derive the term $\mu_r^{i,l}$ in (\ref{eq:mu_r_i}). Notice that with probability $p_{rd}$ node $\mathcal{S}$ conducts a Relay-to-Destination operation, and it will equally likely choose one of the $n-2$ nodes (except $\mathcal{S}$ and $\mathcal{D}$) as its receiver. Thus, when there are $l$ non-empty relay queues in the relay buffer, the corresponding service rate $\mu_r^{i,l}$ is determined as
\begin{equation}
\mu_r^{i,l}=l \cdot \frac{p_{rd}}{n-2}. \label{eq:mu_r_il}
\end{equation}

To determine the conditional probability $P_{l|i}$, we adopt the following occupancy approach proposed in \cite{Stark_BOOK02}. First, for the relay buffer with $i$ packets, where each packet may be destined for any one of the $n-2$ nodes (except $\mathcal{S}$ and $\mathcal{D}$), the number of all possible cases $N_i$ is 
\[\binom{n-3+i}{i}.\]
Then, for the relay buffer with $i$ packets, where these packets are destined for only $l$ different nodes, the number of possible cases $N_{l|i}$ is 
\[\binom{n-2}{l} \cdot \binom{(l-1)+(i-l)}{i-l}.\] 
Finally, since the locations of nodes are independently and uniformly distributed, each case occurs with equal probability. According to the \emph{Classical Probability}, we have
\begin{equation}
P_{l|i}=\frac{N_{l|i}}{N_i}=\frac{\binom{n-2}{l} \cdot \binom{i-1}{i-l}}{\binom{n-3+i}{i}}. \label{eq:P_l|i}
\end{equation}

Substituting (\ref{eq:mu_r_il}) and (\ref{eq:P_l|i}) into (\ref{eq:mu_r_i}), $p_{i,i-1}$ is determined as
\begin{equation*}
p_{i,i-1}=\mu_r^i=p_{rd} \cdot \frac{i}{n-3+i}, \ 1 \leq i \leq B_r.
\end{equation*}

\section{Proof of Corollary~\ref{corollary:OSD_feedback}} \label{appendix:OSD_feedback}

For the network scenario with feedback, although node $\mathcal{S}$ gets the chance to execute the Source-to-Relay operation in a time slot, it still remains idle if the relay buffer of its intended receiver is full (with the overflow probability $\pi_r(B_r)$). Thus, the service rate $\mu_s$ of source buffer of node $\mathcal{S}$ is given by
\begin{equation*}
\mu_s=p_{sd}+p_{sr} \cdot (1-\pi_r(B_r)).
\end{equation*}

Based on the similar analysis as that in Section~\ref{subsection:OSD_nofeedback}, the OSD $\mathbf{\Pi}_s$ of source buffer here can also be determined by expression (\ref{eq:OSD_source}), and the one-step transition probabilities of the birth-death chain of relay buffer can be determined as
\begin{align*}
& p_{i,i+1}=\lambda_r^{\text {\scriptsize +}},  \\
& p_{i,i-1}=p_{rd} \cdot \frac{i}{n-3+i}, \nonumber
\end{align*}
where $\lambda_r^{\text {\scriptsize +}}$ denotes the packet arrival rate of the relay buffer when the relay buffer is not full. Regarding the evaluation of $\lambda_r^{\text{\scriptsize +}}$, we have
\begin{equation}
\lambda_r^{\text {\scriptsize +}}  \cdot  (1 \!-\! \pi_r(B_r)) \!+\! 0  \cdot  \pi_r(B_r) \!=\! (n\!-\!2)\lambda_s^{\text{\small -}}  \cdot  \frac{p_{sr}(1 \!-\! \pi_r(B_r))}{\mu_s(n\!-\!2)}, \label{eq:lambda_r+_FB_1} 
\end{equation}
\begin{equation}
\Rightarrow \lambda_r^{\text{\scriptsize +}}=\lambda_s^{\text{\small -}} \frac{p_{sr}}{\mu_s} = p_{sr} \cdot (1-\pi_s(0)), \label{eq:lambda_r+_FB}
\end{equation}
where $\lambda_s^{\text{\small -}}$ denotes the packet departure rate of a source buffer, and (\ref{eq:lambda_r+_FB}) follows from (\ref{eq:lambda_s-}). Notice that the transition probabilities here are the same as that under the scenario without feedback, thus the OSD $\mathbf{\Pi}_r$ of the relay buffer here can also be determined by expression (\ref{eq:OSD_relay}).

\section{Fixed Point Iteration for Solving $\pi_r(B_r)$} \label{appendix:fixed_point_iteration}
Since $\pi_r(B_r)$ is the fixed point of equation (\ref{eq:self-mapping}), we apply the fixed point iteration to solve $\pi_r(B_r)$. The detailed algorithm of the fixed point iteration is summarized in Algorithm~\ref{algorithm:fixed_point_iteration}, where $\delta$ represents the accuracy can be achieved by the algorithm\footnote{The smaller $\delta$ is, the higher accuracy can be achieved, coming with a cost of more iterations. In our experiment, we set $\delta$ to be $10^{-6}$ to achieve a high accuracy. The execution time of the algorithm under this setting is usually less than $0.2$ seconds.}.

\begin{algorithm}[!ht]
\caption{Fixed Point Iteration}
\label{algorithm:fixed_point_iteration}
\begin{algorithmic}[1]
\REQUIRE ~~\\                          
Basic network parameters $\{n, B_s, B_r, \lambda_s^{\text{\scriptsize +}}, p_{sd}, p_{sr}, p_{rd}\}$;
\ENSURE ~~\\                         
Relay buffer overflow probability $\pi_r(B_r)$;
\STATE Set $x_1=0$ and $i=1$;
\WHILE{($x_i-x_{i-1}\geq \delta) \lor (i=1)$}
  \STATE $i=i+1$;
  \STATE $\mu_s=p_{sd}+p_{sr}\cdot (1-x_{i-1})$;
	\STATE $\tau=\frac{\lambda_s^{\text{\scriptsize +}}(1-\mu_s)}{\mu_s(1-\lambda_s^{\text{\scriptsize +}})}$;
	\STATE $\pi_s(0)=\frac{\mu_s-\lambda_s^{\text{\scriptsize +}}}{\mu_s-\lambda_s^{\text{\scriptsize +}} \cdot \tau^{B_s}}$;
	\STATE $x_i=\frac{\mathrm{C}_{B_r} (1-\pi_s(0))^{B_r}} {\sum\limits_{k=0}^{B_r} \mathrm{C}_k (1-\pi_s(0))^k}$;
\ENDWHILE
\STATE $\pi_r(B_r)=x_i$;
\RETURN $\pi_r(B_r)$;
\end{algorithmic}
\end{algorithm}

\section{Proof of Theorem~\ref{theorem:throughput_delay}} \label{appendix:throughput_delay}

Let $T_1$ and $T_2$ denote the packet delivery rates at the destination of node $\mathcal{S}$ through the one-hop transmission and the two-hop transmission respectively, then we have
\begin{align}
& T_1= \lambda_s^{\text{\small-}} \cdot \frac{p_{sd}}{\mu_s}, \label{eq:T1} \\ 
& T_2= \lambda_s^{\text{\small-}} \cdot \frac{p_{sr}\left(1-\pi_r(B_r) \right)}{\mu_s},   \label{eq:T2}
\end{align}
where $\lambda_s^{\text{\small -}}$ denotes the packet departure rate of source buffer of $\mathcal{S}$. Substituting (\ref{eq:lambda_s-}) into (\ref{eq:T1}) and (\ref{eq:T2}), then (\ref{eq:throughput}) follows from $T=T_1+T_2$.

Regarding the expected E2E delay $\mathbb{E}\{D\}$, we focus on a tagged packet $\mathrm{p}$ of node $\mathcal{S}$ and evaluate its expected source queuing delay $\mathbb{E}\{D_{S_Q}\}$ and expected delivery delay $\mathbb{E}\{D_D\}$, respectively. For the evaluation of $\mathbb{E}\{D_{S_Q}\}$ we have
\begin{equation}
\mathbb{E}\{D_{S_Q}\}=\frac{L_s}{\mu_s}. \label{eq:queuing_delay}
\end{equation}

Let $\pi_s^*(i)$ ($0\leq i\leq B_s-1$) denote the probability that there are $i$ packets in the source buffer conditioned on that the source buffer is not full, then $\pi_s^*(i)$ is determined as \cite{Daduna_BOOK01}
\begin{equation*}
\pi_s^*(i)=\frac{\lambda_s^{\text{\scriptsize +}}}{(1-\lambda_s^{\text{\scriptsize +}})^2}\tau^i \cdot H_1^{-1}, \ 0 \leq i \leq B_r-1
\end{equation*}
where $H_1$ is the normalization constant. Since $\sum\limits_{i=1}^{B_s-1}{\pi_s^*(i)}=1$, we have
\begin{equation*}
\pi_s^*(i)=\frac{1-\tau}{1-\tau^{B_s}}\tau^i, \ 0 \leq i \leq B_r-1.
\end{equation*}
Then $L_s$ is given by
\begin{equation*}
L_s =\sum\limits_{i=0}^{B_s-1}{i\pi_s^*(i)}=\frac{\tau-B_s\tau^{B_s}+(B_s-1)\tau^{B_s+1}}{(1-\tau)(1-\tau^{B_s})}. 
\end{equation*}

After moving to the HoL in its source buffer, packet $\mathrm{p}$ will be sent out by node $\mathcal{S}$ with mean service time $1/\mu_s$, and it may be delivered to its destination directly or forwarded to a relay. Let $\mathbb{E}\{D_R\}$ denote the expected time that $\mathrm{p}$ takes to reach its destination after it is forwarded to a relay, then we have
\begin{equation}
\mathbb{E}\{D_D\} =\frac{1}{\mu_s}+\frac{T_1}{T_1+T2}\cdot 0+\frac{T_2}{T_1+T2}\cdot\mathbb{E}\{D_R\}.  \label{eq:delivery_delay} 
\end{equation}

Based on the OSD $\mathbf{\Pi}_r$, $L_r$ is given by (\ref{eq:relay_length}). Due to the symmetry of relay queues in a relay buffer, the mean number of packets in one relay queue is $L_r/(n-2)$, and the service rate of each relay queue is $p_{rd}/(n-2)$. Thus, $\mathbb{E}\{D_R\}$ can be determined as 
\begin{equation}
\mathbb{E}\{D_R\}=\left(\frac{L_r}{n-2}+1 \right) \cdot \left(\frac{p_{rd}}{n-2} \right)^{-1}. \label{eq:D_R}
\end{equation}
Substituting (\ref{eq:D_R}) into (\ref{eq:delivery_delay}), then (\ref{eq:E2E_delay}) follows from $\mathbb{E}\{D\}=\mathbb{E}\{D_{S_Q}\}+\mathbb{E}\{D_D\}$.

\section{Proof of Corollary~\ref{corollary:feedback}} \label{appendix:feedback}

From expressions (\ref{eq:mu_s_NF}) and (\ref{eq:mu_s_FB}), we can see that the for a given packet generating rate $\lambda_s^{\text{\scriptsize +}}$, the service rate $\mu_s$ of the source buffer under the scenario with feedback is smaller than that under the scenario without feedback. From (\ref{eq:OSD_source}) we have
\begin{align}
\frac{\partial \pi_s(0)}{\partial \mu_s} & \! = \!\frac{\mu_s\!-\!\lambda_s^{\text{\scriptsize +}}\tau^{B_s}\!-\!\left(1\!-\!\lambda_s^{\text{\scriptsize +}}B_s\tau^{B_s-1}\frac{\partial \tau}{\partial \mu_s}\right)(\mu_s\!-\!\lambda_s^{\text{\scriptsize +}})}{(\mu_s-\lambda_s^{\text{\scriptsize +}}\tau^{B_s})^2} \nonumber \\
& \!=\! \frac{\lambda_s^{\text{\scriptsize +}}-\lambda_s^{\text{\scriptsize +}}\tau^{B_s}-B_s\frac{\lambda_s^{\text{\scriptsize +}}(\mu_s-\lambda_s^{\text{\scriptsize +}})}{\mu_s(1-\mu_s)}\tau^{B_s}}{(\mu_s-\lambda_s^{\text{\scriptsize +}}\tau^{B_s})^2} \nonumber \\
& \!=\!\frac{\lambda_s^{\text{\scriptsize +}}(\mu_s-\lambda_s^{\text{\scriptsize +}})^2}{(\mu_s\!-\!\lambda_s^{\text{\scriptsize +}}\tau^{B_s})^2 \!\cdot\! \mu_s^2 \!\cdot\! (1\!-\!\lambda_s^{\text{\scriptsize +}})  }\!\cdot\! \sum\limits_{i=0}^{B_s-1}{\left(1+\frac{i}{1-\mu_s}\right)\tau^i} \nonumber \\
& \!>\!0, \label{eq:pi_s_0_mu_s}
\end{align}
which indicates that $\pi_s(0)$ under the scenario with feedback is smaller than that under the scenario without feedback.

We let $r=\frac{1}{1-\pi_s(0)}$ and substitute $r$ into (\ref{eq:throughput}), then $T$ can be expressed as 
\begin{equation}
T = p_{sd} \cdot \frac{1}{r} +p_{sr} \cdot \frac{1}{g(r)}, \label{eq:T_r}
\end{equation}
where $g(r)=r \cdot \left(1+ \frac{\mathrm{C}_{B_r} } {h(r)}\right)$ and $h(r)=\sum\limits_{i=0}^{B_r-1}{\mathrm{C}_i r^{B_r-i}}$. Regarding the derivative of $g(r)$ we have
\begin{equation}
g'(r) =\frac{1}{h(r)^2} \underbrace{\left\{h(r)(h(r)+\mathrm{C}_{B_r})-r\mathrm{C}_{B_r}h'(r)\right\}}_{(a)}, 
\end{equation}
where
\begin{align}
(a) & = \sum\limits_{i=0}^{B_r-1}{\mathrm{C}_i r^{B_r-i}} \cdot \sum\limits_{i=0}^{B_r-1}{\mathrm{C}_i r^{B_r-i}} \nonumber \\
    & - \mathrm{C}_{B_r}\sum\limits_{i=0}^{B_r-1}(B_r-i)\mathrm{C}_i r^{B_r-i} \nonumber \\
		& = \sum\limits_{i=1}^{B_r}{\mathrm{C}_{B_r-i} r^i} \cdot \sum\limits_{i=0}^{B_r}{\mathrm{C}_{B_r-i} r^i}-\sum\limits_{i=1}^{B_r} i \mathrm{C}_{B_r}\mathrm{C}_{B_r-i} r^i \nonumber \\
		& = \sum\limits_{i=1}^{B_r} \left(\sum\limits_{j=0}^{i-1}{\mathrm{C}_{B_r-j}r^j \mathrm{C}_{B_r-i+j}r^{i-j}}-i\mathrm{C}_{B_r}\mathrm{C}_{B_r-i}r^i \right) \nonumber \\
		& + \sum\limits_{i=B_r+1}^{2B_r} \sum\limits_{j=i-B_r}^{B_r} \mathrm{C}_{B-j}r^j \mathrm{C}_{B-i+j}r^{i-j} \nonumber \\
		& > \sum\limits_{i=1}^{B_r} {\left(\sum\limits_{j=0}^{i-1}{\mathrm{C}_{B_r-j}\mathrm{C}_{B_r-i+j}}\!-\!i\mathrm{C}_{B_r}\mathrm{C}_{B_r-i}\right) r^i} \!>\!0, \label{eq:larger_0}
\end{align}
here (\ref{eq:larger_0}) is because that $\mathrm{C}_{B_r-j}\mathrm{C}_{B_r-i+j} > \mathrm{C}_{B_r}\mathrm{C}_{B_r-i}$ for $0<j<i$.

We can see from (\ref{eq:pi_s_0_mu_s}) that $\pi_s(0)$ increases as $\mu_s$ increases, and from (\ref{eq:T_r})$-$(\ref{eq:larger_0}) that $T$ increases as $\pi_s(0)$ decreases. Thus, we can conclude that $T$ under the scenario with feedback is larger than that under the scenario without feedback, which indicates that adopting the feedback mechanism improves the throughput performance.

\section{Proof of Lemma~\ref{lemma:as_lambda_increase}} \label{appendix:as_lambda_increase}

For the scenario without feedback, we know from (\ref{eq:OSD_source}) that
\begin{align}
\frac{\partial \pi_s(0)}{\partial \lambda_s^{\text{\scriptsize +}}} & \! = \!\frac{-\mu_s\!+\!\lambda_s^{\text{\scriptsize +}}\tau^{B_s}\!+\!\left(\tau^{B_s}\!+\!\lambda_s^{\text{\scriptsize +}}B_s\tau^{B_s-1}\frac{\partial \tau}{\partial \lambda_s^{\text{\scriptsize +}}}\right)(\mu_s\!-\!\lambda_s^{\text{\scriptsize +}})}{(\mu_s-\lambda_s^{\text{\scriptsize +}}\tau^{B_s})^2} \nonumber \\
& \!=\! \frac{-\mu_s+\mu_s\tau^{B_s}+B_s\frac{\mu_s-\lambda_s^{\text{\scriptsize +}}}{1-\lambda_s^{\text{\scriptsize +}}}\tau^{B_s}}{(\mu_s-\lambda_s^{\text{\scriptsize +}}\tau^{B_s})^2} \nonumber \\
& \!=\!\frac{-(\lambda_s^{\text{\scriptsize +}}-\mu_s)^2}{(\mu_s\!-\!\lambda_s^{\text{\scriptsize +}}\tau^{B_s})^2 \!\cdot\! (1\!-\!\lambda_s^{\text{\scriptsize +}})^2 \!\cdot\! \mu_s}\!\cdot\! \sum\limits_{i=1}^{B_s}{i\tau^{i-1}} \!<\!0. \label{eq:pi_s_0_lambda_s}
\end{align}
Thus, as $\lambda_s^{\text{\scriptsize +}}$ increases, $\pi_s(0)$ decreases which leads to an increase in $T$ (refer to the analysis in Appendix~\ref{appendix:feedback}).

For the scenario with feedback, as $\lambda_s^{\text{\scriptsize +}}$ increases, the MANET tends to be more congested with a larger $\pi_r(Br)$. Thus, we know from (\ref{eq:mu_s_FB}) that the corresponding $\mu_s$ decreases, and then from (\ref{eq:pi_s_0_mu_s}) that $\pi_s(0)$ decreases, leading to an increase in $T$.


\section{Proof of Corollary~\ref{corollary:buffer_infinite}} \label{appendix:buffer_infinite}
From an intuitive point of view, a larger buffer implies that more packets can be stored and packet loss can be reduced, thus a higher throughput can be achieved. More formally, from (\ref{eq:OSD_source}) we have
\begin{align}
&\pi_s(0)|_{B_s=K+1}-\pi_s(0)|_{B_s=K} \nonumber \\ 
 = & \frac{\lambda_s^{\text{\scriptsize +}}\tau^{K}(\mu_s-\lambda_s^{\text{\scriptsize +}})(\tau-1)}{(\mu_s-\lambda_s^{\text{\scriptsize +}}\tau^{K+1})(\mu_s-\lambda_s^{\text{\scriptsize +}}\tau^{K})} <0, \label{eq:derivative_B_s}
\end{align}
where (\ref{eq:derivative_B_s}) follows since $\tau>1$ when $\lambda_s^{\text{\scriptsize +}}>\mu_s$ and $\tau<1$ when $\lambda_s^{\text{\scriptsize +}}<\mu_s$. Then we can conclude that as $B_s$ increases, $\pi_s(0)$ decreases, leading to an increase in $T$.

Let $r=\frac{1}{1-\pi_s(0)}$ and substitute $r$ into (\ref{eq:OSD_relay}), then we have 
\begin{align*}
  & \pi_r(B_r)|_{B_r=K+1}-\pi_r(B_r)|_{B_r=K} \nonumber \\
= & \frac{\mathrm{C}_{K+1}r^{-K-1}}{\sum\limits_{i=0}^{K+1}\mathrm{C}_i r^{-i}}-\frac{\mathrm{C}_K r^{-K}}{\sum\limits_{i=0}^K\mathrm{C}_i r^{-i}} \\
= & \frac{\underbrace{\mathrm{C}_{K+1} r^{-K-1}\sum\limits_{i=0}^K\mathrm{C}_i r^{-i}-\mathrm{C}_K r^{-K}\sum\limits_{i=0}^{K+1}\mathrm{C}_i r^{-i}}_{(b)}}{\sum\limits_{i=0}^{K+1}\mathrm{C}_i r^{-i} \cdot \sum\limits_{i=0}^K\mathrm{C}_i r^{-i}},
\end{align*}
where 
\begin{align*}
(b)&\!=\!\mathrm{C}_{K\!+\!1} r^{\!-\!K\!-\!1}\sum\limits_{i=0}^K\mathrm{C}_i r^{\!-\!i}\!-\!\mathrm{C}_K r^{\!-\!K}\sum\limits_{i=1}^{K\!+\!1}\mathrm{C}_i r^{\!-\!i}\!-\!\mathrm{C}_K r^{\!-\!K} \\
   &\!<\! \sum\limits_{i=0}^K {\left(\mathrm{C}_{K+1}\mathrm{C}_i-\mathrm{C}_K\mathrm{C}_{i+1} \right) r^{-k-i-1}} <0.
\end{align*}
Then we can conclude that as $B_r$ increases, $\pi_r(B_r)$ decreases, leading to an increase in $T$ (refer to expression (\ref{eq:throughput})). 

Regarding the infinite source buffer (i.e., $B_s\to\infty$), $\tau \geq 1$ when $\lambda_s^{\text{\scriptsize +}} \geq \mu_s$, and we have
\begin{align*}
& \lim\limits_{B_s\to\infty} \pi_s(0)=\lim\limits_{B_s\to\infty}\frac{\mu_s-\lambda_s^{\text{\scriptsize +}}}{\mu_s-\lambda_s^{\text{\scriptsize +}}\tau^{B_s}}=0,  \\
& \begin{aligned} 
  \lim\limits_{B_s\to\infty} T & =p_{sd} + p_{sr}\cdot\frac{\sum\limits_{k=1}^{B_r} \mathrm{C}_{k-1}}{\sum\limits_{k=0}^{B_r} \mathrm{C}_k ^k} \\
	                             & =p_{sd}+p_{sr} \frac{B_r}{n-2+B_r}=T_c.
	\end{aligned}
\end{align*}
According to the Queuing theory \cite{Daduna_BOOK01}, for a Bernoulli/Bernoulli queue (i.e., the buffer size is infinite), its queue length tends to infinity when the corresponding arrival rate is equal to or larger than the service rate. Thus, we have $L_s \to \infty$, which leads that $\mathbb{E}\{D_{S_Q}\} \to \infty$ and $\mathbb{E}\{D\} \to \infty$.

When $\lambda_s^{\text{\scriptsize +}}<\mu_s$, $\tau<1$, and we have
\begin{align*}
&\lim\limits_{B_s\to\infty} \pi_s(0)=\lim\limits_{B_s\to\infty}\frac{\mu_s-\lambda_s^{\text{\scriptsize +}}}{\mu_s-\lambda_s^{\text{\scriptsize +}}\tau^{B_s}}=1-\frac{\lambda_s^{\text{\scriptsize +}}}{\mu_s}, \\
&\lim\limits_{B_s\to\infty} T=p_{sd}\cdot \frac{\lambda_s^{\text{\scriptsize +}}}{\mu_s}+p_{sr}\cdot\frac{\sum\limits_{k=0}^{B_r-1} \mathrm{C}_k (\frac{\lambda_s^{\text{\scriptsize +}}}{\mu_s})^{k+1}}{\sum\limits_{k=0}^{B_r} \mathrm{C}_k (\frac{\lambda_s^{\text{\scriptsize +}}}{\mu_s})^k}.
\end{align*}
Based on the analysis in Appendix~\ref{appendix:throughput_delay}, $L_s$ is determined as
\begin{equation}
\lim\limits_{B_s\to\infty}L_s=\lim\limits_{B_s\to\infty}\frac{1-\tau}{1-\tau^{B_s}}\sum\limits_{i=0}^{B_s-1}{i\tau^i}=\frac{\tau}{1-\tau}. \label{eq:Ls_Bs_infinite}
\end{equation}
Substituting (\ref{eq:Ls_Bs_infinite}) into (\ref{eq:E2E_delay}) we obtain (\ref{eq:D_Bs_infinite}).

Regarding the infinite relay buffer (i.e., $B_r\to\infty$), from (\ref{eq:OSD_relay}) and (\ref{eq:relay_length}) we have
\begin{align}
\lim\limits_{B_r\to\infty}\pi_r(B_r) & =\lim\limits_{B_r\to\infty} \mathrm{C}_{B_r}(1-\pi_s(0))^{B_r}\cdot \pi_s(0)^{n-2}  \label{eq:taylor_expansion} \\
                                     & \leq \lim\limits_{B_r\to\infty} (B_r+n)^n (1-\pi_s(0))^{B_r} \nonumber \\
																		 & \leq \lim\limits_{B_r\to\infty} 2^n B_r^n (1-\pi_s(0))^{B_r} \nonumber \\
																		 & =\lim\limits_{B_r\to\infty} \frac{2^n n!(1-\pi_s(0))^{B_r}}{(\ln{\frac{1}{1-\pi_s(0)}})^n}=0, \label{eq:pi_r_infinite}
\end{align}
\begin{align}
\lim\limits_{B_r\to\infty} L_r & =\frac{\sum\limits_{k \geq 0}{k\mathrm{C}_k(1-\pi_s(0))^k}}{\sum\limits_{k \geq 0}{\mathrm{C}_k(1-\pi_s(0))^k}} \nonumber \\
                 & = \frac{-(1-\pi_s(0))\cdot \left(\sum\limits_{k \geq 0}{\mathrm{C}_k(1-\pi_s(0))^k}\right)'}{\sum\limits_{k \geq 0}{\mathrm{C}_k(1-\pi_s(0))^k}} \nonumber \\
                 & = -(1-\pi_s(0))\cdot \left(\pi_s(0)^{2-n}\right)' \cdot  \pi_s(0)^{n-2}                   \label{eq:taylor_expansion1} \\
								 & =\frac{(n-2)(1-\pi_s(0))}{\pi_s(0)}, \label{eq:Lr_Br_infinite}  
\end{align}
where (\ref{eq:taylor_expansion}) and (\ref{eq:taylor_expansion1}) follow since $\sum\limits_{k\geq 0}\mathrm{C}_k(1-\pi_s(0))^k$ is just the Taylor-series expansion \cite{Stewart_BOOK11} of $\pi_s(0)^{2-n}$, and (\ref{eq:pi_r_infinite}) follows from the L'H{\^o}pital's rule \cite{Stewart_BOOK11}. Substituting (\ref{eq:pi_r_infinite}) into (\ref{eq:throughput}) we obtain (\ref{eq:T_Br_infinite}), and substituting (\ref{eq:pi_r_infinite}) and (\ref{eq:Lr_Br_infinite}) into (\ref{eq:E2E_delay}) we obtain (\ref{eq:D_Br_infinite}).

Regarding the MANET without buffer constraint (i.e., $B_s\to\infty$ and $B_r\to\infty$), we can directly obtain (\ref{eq:T_Bs_Br_infinite}) and (\ref{eq:D_Bs_Br_infinite}) by combining the corresponding results of the infinite source buffer scenario and the infinite relay buffer scenario.


\section{Derivations of Probabilities $p_{sd}$, $p_{sr}$ and $p_{rd}$} \label{appendix:basic_probabilities}
For a cell-partitioned MANET with LS-MAC, the event that node $\mathcal{S}$ gets the chance to execute the Source-to-Destination (resp. Source-to-Relay or Relay-to-Destination) operation in a time slot can be divided into the following sub-events: (1) its destination is (resp. is not) in the same cell with $\mathcal{S}$; (2) other $k$ out of $n-2$ nodes are in the same cell with $\mathcal{S}$, while the remaining $n-2-k$ nodes are not in this cell; (3) $\mathcal{S}$ contends for the wireless channel access successfully. Thus we have
\begin{align*}
p_{sd} & =\sum_{k=0}^{n-2}{\binom{n-2}{k} (\frac{1}{m^2})^{k+1} (1-\frac{1}{m^2})^{n-2-k}  \cdot \frac{1}{k+2}}   \\ 
       & =\sum_{k=0}^{n-2}{\binom{n-1}{k+1} (\frac{1}{m^2})^{k+1} (1-\frac{1}{m^2})^{n-2-k}  \cdot \frac{1}{k+2}} \\ 
			 & -\sum_{k=0}^{n-3}{\binom{n-2}{k+1} (\frac{1}{m^2})^{k+1} (1-\frac{1}{m^2})^{n-2-k}  \cdot \frac{1}{k+2}} \\
			 & =\frac{m^2}{n}\left \{ 1-(1-\frac{1}{m^2})^n \right \}-(1-\frac{1}{m^2})^{n-1}                           \\ 
			 & -\frac{m^2-1}{n-1}\left \{ 1-(1-\frac{1}{m^2})^{n-1} \right \} +(1-\frac{1}{m^2})^{n-1}                  \\ 
			 & =\frac{m^2}{n}-\frac{m^2-1}{n-1}+(\frac{m^2-1}{n-1}-\frac{m^2-1}{n})(1-\frac{1}{m^2})^{n-1}, \nonumber
\end{align*}
and
\begin{align*}
p_{sr} & =p_{rd}  \\
       & =\frac{1}{2}\sum_{k=1}^{n-2}{\binom{n-2}{k} (\frac{1}{m^2})^k (1-\frac{1}{m^2})^{n-1-k} \cdot \frac{1}{k+1}}  \\
       & =\frac{1}{2}\left \{ \frac{m^2-1}{n-1}-\frac{m^2}{n-1} (1-\frac{1}{m^2})^n-(1-\frac{1}{m^2})^{n-1} \right \}  
\end{align*}

For a cell-partitioned MANET with EC-MAC, by applying the similar approach and algebraic operations we have
\begin{align*}
p_{sd} & = \frac{1}{\varepsilon ^2} \left \{ \sum_{k=0}^{n-2} {\binom{n-2}{k}(\frac{1}{m^2})^{k+1}(1-\frac{1}{m^2})^{n-2-k} \cdot \frac{1}{k+2}} \right.  \\
       & + \left. \sum_{k=0}^{n-2} {\binom{n-2}{k} (\frac{1}{m^2})^{k+1} (1-\frac{1}{m^2})^{n-2-k}  \cdot \frac{4v^2-4v}{k+1} } \right \}  \\
			 & = \frac{1}{\varepsilon^2} \left \{ \frac{\Gamma-\frac{m^2}{n}}{n-1} + \frac{m^2-1-(\Gamma-1)n}{n(n-1)} (1-\frac{1}{m^2}) ^{n-1}  \right \}, 
\end{align*}
and
\begin{align*} 
p_{sr} &=p_{rd}  \\
       &= \frac{1}{2 \varepsilon^2} \frac{m^2-\Gamma}{m^2} \cdot \\
       &  \left \{ \sum_{k=1}^{n-2}{ \binom{n-2}{k} (\frac{1}{m^2})^k (1-\frac{1}{m^2})^{n-2-k} \cdot \frac{1}{k+1}}\right.  \\
       & + \left. \sum_{k=1}^{n-2} { \binom{n-2}{k} (\frac{\Gamma-1}{m^2})^k (\frac{m^2-\Gamma}{m^2})^{n-2-k} } \right\}  \\
			 & = \frac{1}{2 \varepsilon^2} \left \{ \frac{m^2-\Gamma}{n-1} (1-(1-\frac{1}{m^2})^{n-1}) - (1-\frac{\Gamma}{m^2})^{n-1}     \right \} .
\end{align*}

\ifCLASSOPTIONcaptionsoff
  \newpage
\fi


\end{document}